\documentclass{article}

\usepackage{amsmath}
\usepackage{pdfsync}
\usepackage{graphicx}

\usepackage{setspace}
\usepackage{authblk}
\usepackage{caption}
\usepackage{subcaption}
\usepackage{subcaption}
\usepackage{float}
\usepackage{multicol}

\usepackage{amssymb,amsfonts,amsmath}
\usepackage{nicefrac}
\usepackage[rightcaption]{sidecap}
\usepackage{color}
\usepackage{ulem} % for \sout

\advance \topmargin by -\headheight
\advance \topmargin by -\headsep

\evensidemargin \oddsidemargin
\title{Heisenberg-limited Rabi spectroscopy}
\author{Ravid Shaniv$^1$, Tom Manovitz$^1$, Yotam Shapira$^1$, Nitzan Akerman$^1$ \& Roee Ozeri$^1$}

\affil{Department of Physics of Complex systems, Weizmann Institute of Science, Rehovot 7610001, Israel
}

\begin{document}

\maketitle

\begin{abstract}
The use of entangled states was shown to improve the fundamental limits of spectroscopy to beyond the standard-quantum limit. In these Heisenberg-limited protocols the phase between two states in an entangled superposition evolves N-fold faster than in the uncorrelated case, where N for example can be the number of entangled atoms in a Greenberger-Horne-Zeilinger (GHZ) state. Here we propose and demonstrate the use of correlated spin-Hamiltonians for the realization of Heisenberg-limited Rabi-type spectroscopy. Rather than probing the free evolution of the phase of an entangled state with respect to a local oscillator (LO), we probe the evolution of an, initially separable, two-atom register under an Ising spin-Hamiltonian with a transverse field. The resulting correlated spin-rotation spectrum is twice as narrow as compared with uncorrelated rotation. We implement this Heisenberg-limited Rabi spectroscopy scheme on the optical-clock electric-quadrupole transition of $^{88}$Sr$^+$ using a two-ion crystal. We further show that depending on the initial state, correlated rotation can occur in two orthogonal sub-spaces of the full Hilbert space, yielding Heisenberg-limited spectroscopy of either the average transition frequency of the two ions or their difference from the mean frequency. The potential improvement of clock stability due to the use of entangled states depends on the details of the method used and the dominating decoherence mechanism. The use of correlated spin-rotations can therefore potentially lead to new paths for clock stability improvement.
\end{abstract}

%Entangled states in quantum metrology
Different quantum technologies rely on entangled states as their primary resource. In quantum metrology it was shown that entangled states can be used to reduce the uncertainty in the spectroscopy of two-level systems (pseudo-spins). The use of spin-squeezed states was shown to reduce the spectroscopy uncertainty in atomic ensembles to below the Standard Quantum Limit (SQL) \cite{Wineland1994, Kasevich1997,  Oberthaler2010, Polzik2010}. In the extreme case of fully entangled spins, it was shown that using Ramsey-like spectroscopy of an N-atom GHZ state, the phase of this state with respect to a LO evolves N-fold faster, leading to Heisenberg limited estimation of the transition frequency \cite{Bollinger1996, Wineland2004, Blatt2011}.

%Ramsey vs. Rabi
In Ramsey spectroscopy the phase of a superposition in free-evolution is compared to a LO. An alternative to Ramsey spectroscopy is Rabi-type spectroscopy, in which the evolution of a state under a time-dependent Hamiltonian is investigated. Under the rotating-wave approximation the Rabi Hamiltonian, $H=\hbar\left[\Omega\sigma_{y} + \delta\sigma_{z}\right]$,  generates spin rotations. Here, $\sigma_{i}$ are the Pauli spin operators, $\Omega$ is the Rabi frequency and $\delta$ is the detuning of the Rabi Hamiltonian from the atomic transition frequency. The initial state can be thought of as a superposition of the Rabi-Hamiltonian dressed-states \cite{CCT1992}. The $\delta=0$ point at which the Rabi Hamiltonian frequency is on-resonance is determined by the point at which spin rotation is maximal. The width of the Rabi spectrum is determined by the gap between the two dressed-states; namely the Rabi frequency $\Omega$.

%Many-body Rabi and Quantum simulations
One can therefore ask whether it is possible to generate Heisenberg-limited Rabi spectroscopy by acting on multi-spin registers with time-dependent Hamiltonians. The need for entanglement in Ramsey Heisenberg-limited spectroscopy suggests that the necessary time-dependent Hamiltonians are many-body interacting Hamiltonians. The simulation of many-body quantum spin-Hamiltonians is a field of growing experimental interest. The adiabatic evolution of ground-states as well as the dynamics of spin-defects under quenching were studied using these synthesized Hamiltonians \cite{Tobias2008, Monroe2011, Greiner2011, Blatt2014}. The application of correlated many-body Hamiltonians typically results in entanglement.

% Rotations in metrological subspaces
Heisenberg limited Ramsey spectroscopy investigates the free-evolution of superpositions in entangled subspaces. By the same token, Heisenberg limited Rabi spectroscopy can be engineered by investigating rotations of states in these entangled subspaces by many-body spin Hamiltonians \cite{OzeriQEC2013}. Similarly to single spin Rabi spectroscopy, the resonance frequency will be determined by the maximal rotation angle and the width of the resonance will be given by the gap between the two eigenstates of the spin-interaction Hamiltonian in this subspace.

% What we did here
In this work we show that an Ising spin-interaction Hamiltonian with a transverse field generates rotations in two orthogonal subspaces of a two-spin Hilbert space. In the spin symmetric subspace, spanned by $\left|\uparrow\uparrow\right\rangle$ and  $\left|\downarrow\downarrow\right\rangle$ this Hamiltonian results in Heisenberg limited Rabi spectroscopy of the average spin transition frequency whereas in the anti-symmetric subspace spanned by $\left|\downarrow\uparrow\right\rangle$ and  $\left|\uparrow\downarrow\right\rangle$ Heisenberg limited Rabi spectroscopy of the spins frequency-difference from the mean transition frequency is performed. We implement this protocol using the optical clock electric-quadrupole transition in a two $^{88}$Sr$^+$ ion-crystal and show that  the resulting correlated spin rotation spectra are indeed twice as narrow as compared with single ion Rabi spectra.

% limits to Heisenberg approaches
Heisenberg limited spectroscopy was shown to have limited value in the improvement of spectroscopic precision after long averaging times. The reason is that, as the sensitivity to the resonance frequency increases, so does the sensitivity to noise and dephasing rates increase \cite{Huelga1997}. In several theoretical investigations it was shown that an improvement of measurement precision or clock stability is possible, however it depends on the exact details of the noise and the spectroscopic method used \cite{Guta2012,Schmidt2017}. Hence, the development of new Heisenberg-limited spectroscopy techniques has the potential of introducing further clock stability improvement under different conditions.

%%%%%%%%%%%%%%%%%%%%%%%%%%%%%%%%%%%%%%%%%%%%%%%%%%%%%%%%%%%%%%%%%
% Beginning theory
%%%%%%%%%%%%%%%%%%%%%%%%%%%%%%%%%%%%%%%%%%%%%%%%%%%%%%%%%%%%%%%%%

We investigate a system of two interacting spins, under the influence of an Ising two-spin Hamiltonian with a transverse field,
\begin{equation}
\begin{split}
H=\hbar\left[\Omega\sigma_{y}\otimes\sigma_{y}+\delta_{1}\left(\sigma_{z}\otimes I+I\otimes\sigma_{z}\right)+\delta_{2}\left(\sigma_{z}\otimes I-I\otimes\sigma_{z}\right)\right].
\end{split}\label{Ising}
\end{equation}
 Here the $\delta_{1}\left(\sigma_{z}\otimes I+I\otimes\sigma_{z}\right)$ term represent magnetic field along the z axis common to both spins and the term $\delta_{2}\left(\sigma_{z}\otimes I-I\otimes\sigma_{z}\right)$ represents the difference between the fields on each spin. The $ \Omega\sigma_{y}\otimes\sigma_{y}$ term is an Ising-type interaction which creates a correlated rotation of the two spins.\\
%%%%%%%%%%%%%%%%%%%%%%%%%%%%%%%%%%%%%%%%%%%%%%%%%%%%%%%%%%%%%%%%%
% continue with theory - scanning and scan result
%%%%%%%%%%%%%%%%%%%%%%%%%%%%%%%%%%%%%%%%%%%%%%%%%%%%%%%%%%%%%%%%%

The Hamiltonian in Eq. \ref{Ising} commutes with $\sigma_{z}\otimes\sigma_{z}$ and therefore conserves state parity and does not mix between the even $\left\{ \left|\downarrow\downarrow\right\rangle ,\left|\uparrow\uparrow\right\rangle \right\} $ and odd $\left\{ \left|\downarrow\uparrow\right\rangle ,\left|\uparrow\downarrow\right\rangle \right\}$ parity subspaces. In addition, the even and odd subspaces are degenerate under the operation of  $\sigma_{z}\otimes I-I\otimes\sigma_{z}$ and $\sigma_{z}\otimes I+I\otimes\sigma_{z}$ respectively. As a result, superpositions of the states $\left|\uparrow\uparrow\right\rangle ,\left|\downarrow\downarrow\right\rangle$ ($\left|\downarrow\uparrow\right\rangle ,\left|\uparrow\downarrow\right\rangle$) are invariant to changes in $\delta_{2}$ ($\delta_{1}$).
%%%%%%%%%%%%%%%%%%%%%%%%%%%%%%%%%%%%%%%%%%%%%%%%%%%%%%%%%%%%%%%%%
% scanning results of the Hamiltonian and Heisenberg limited spectroscopy
%%%%%%%%%%%%%%%%%%%%%%%%%%%%%%%%%%%%%%%%%%%%%%%%%%%%%%%%%%%%%%%%%
The two subspaces above can be thought of as two super-spin-half metrological subspaces. As an example in the even subspace the two basis states of a super-spin-half system with $\left|\tilde{\uparrow}\right\rangle :=\left|\uparrow\uparrow\right\rangle ,\left|\tilde{\downarrow}\right\rangle :=\left|\downarrow\downarrow\right\rangle $. Ising spin coupling acts as a $y$ rotation in this subspace, $\tilde{\sigma}_{y}=\sigma_{y}\otimes\sigma_{y}$ and a $z$ rotation is generated by $\tilde{\sigma}_{z}=\delta_{1}\left(\sigma_{z}\otimes I+I\otimes\sigma_{z}\right)$. Rotations around $z$ in the anti-symmetric subspace are generated by $\delta_{2}\left(\sigma_{z}\otimes I-I\otimes\sigma_{z}\right)$. The Ising Hamiltonian in Eq. \ref{Ising} therefore performs Rabi spectroscopy in the two spin subspaces with a Rabi frequency $\Omega$ and a detuning $2\delta_1$ or $2\delta_2$ respectively. The factor of two in the detuning results in a two-fold narrowing of the Rabi resonance, leading to Heisenberg limited determination of the resonance frequency under spin projection noise. Notice that a general two-spin state is a direct sum of states in these two subspaces. A measurement will therefore lead to a single bit of spectroscopic information, thus increasing the standard deviation due to projection noise.

 \begin{figure}[H]
\centering
\includegraphics[width=7cm]{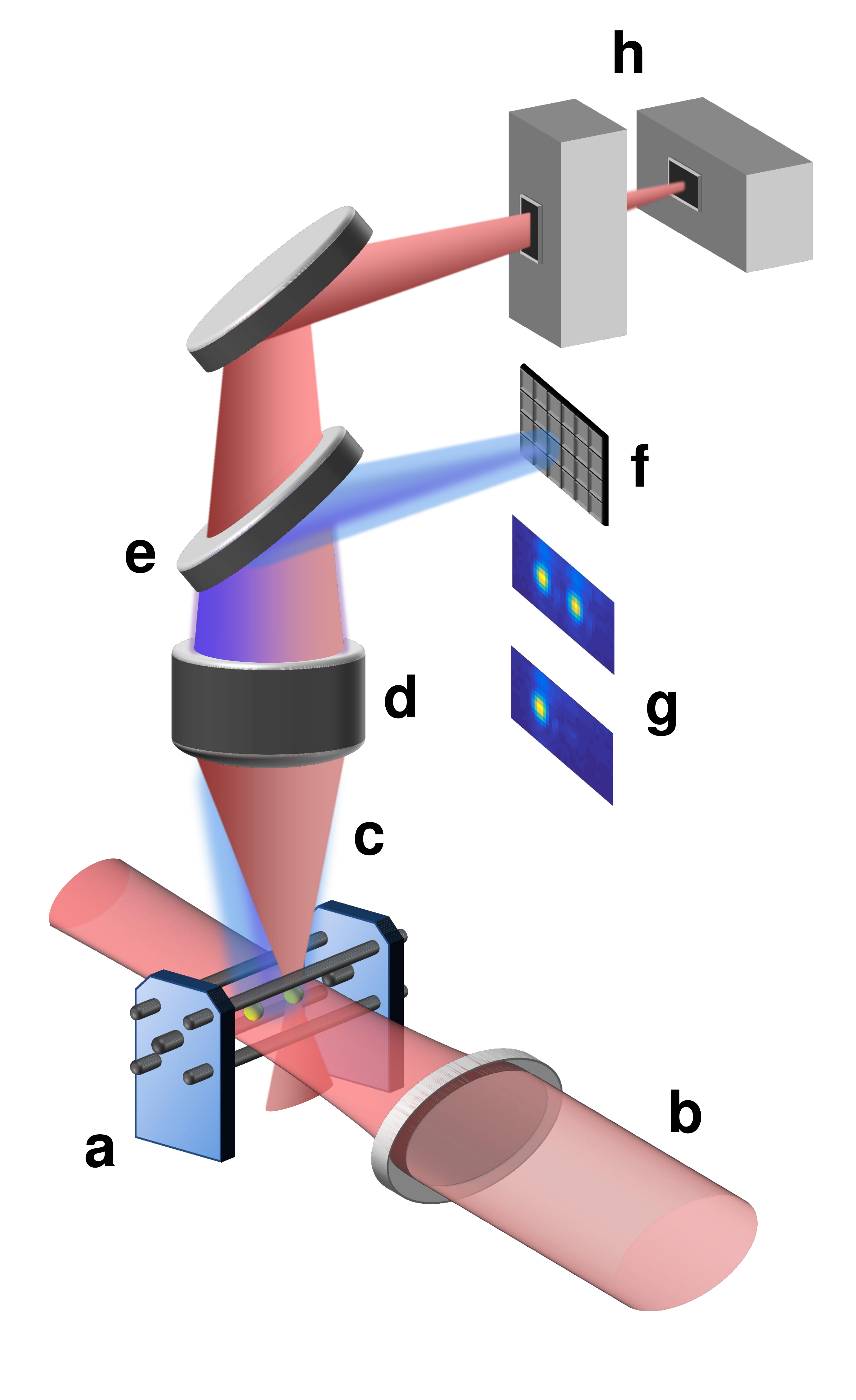}
\caption{\textbf{Schematic layout of our experiment.} Two ions trapped in a linear Paul trap \textbf{(a)} are addressed with two clock-transition laser beams. A global beam addresses both ions equally \textbf{(b)} and a tightly focused single-ion addressing beam can be tuned to address ions individually \textbf{(c)}. A magnetic field of 3.1 G is aligned along the global beam propagation direction (not marked in the figure). An objective lens \textbf{(d)} focuses the single-ion addressing beam onto the ion-crystal. The same objective lens collects  fluorescence from the ions at 422 nm used for state readout. The two wavelengths are separated with a dichroic mirror \textbf{(e)}. The fluorescence light is focused on an EMCCD camera \textbf{(f)} camera images of two and a single fluorescing ions are shown \textbf{(g)}. The single-ion addressing beam passes through a double AOM system in a XY configuration, allowing fast 2D scan of the beam's position and fast shifting of the beam from one ion to the another \textbf{(h)}.}
\end{figure}

%%%%%%%%%%%%%%%%%%%%%%%%%%%%%%%%%%%%%%%%%%%%%%%%%%%%%%%%%%%%%%%%%
% Experimental system
%%%%%%%%%%%%%%%%%%%%%%%%%%%%%%%%%%%%%%%%%%%%%%%%%%%%%%%%%%%%%%%%%

In our experiment, the pseudo-spin states are the two optical-clock transition levels, $5S_{j=\frac{1}{2},m_{j}=-\frac{1}{2}}$ and $4D_{j=\frac{5}{2},m_{j}=-\frac{3}{2}}$, in trapped $^{88}Sr^{+}$ ions. Our ions are trapped in a linear Paul trap and laser-cooled to the ground state of motion \cite{WinelandComprehensive1998,NitzanApplied2012,Nitzan2015} in the axial direction. We drive the optical clock transition using a 674 $\mathrm{nm}$ narrow linewidth ($<50$ Hz) laser. We address the two ion crystal with a single large-waist beam which implements both global rotations as well as the transverse Ising Hamiltonian. Alternatively we individually address a single ion of our choice using a tightly focused laser beam. The state of our ion is detected using state-selective fluorescence detection. An illustration of our experimental setup is shown in Fig. 1. More details about it can be found in \cite{NitzanApplied2012,Nitzan2015,TomThesis2016}. \\

%%%%%%%%%%%%%%%%%%%%%%%%%%%%%%%%%%%%%%%%%%%%%%%%%%%%%%%%%%%%%%%%%
% Hamiltonian using Sorensen molmer
%%%%%%%%%%%%%%%%%%%%%%%%%%%%%%%%%%%%%%%%%%%%%%%%%%%%%%%%%%%%%%%%%

The Ising Hamiltonian in Eq. \ref{Ising} is realized using a M\o lmer-S\o rensen- (MS) interaction \cite{MS_Gate2000}. We denote the clock transition carrier frequency of ion 1 and ion 2 as $\omega_{0}^1$ and $\omega_{0}^2$ respectively. The ions are illuminated with a bichromatic 674 $\mathrm{nm}$ laser beam at frequencies
\begin{equation}
\omega_{\pm}=\omega_{0}\pm\nu\pm\varepsilon-\delta \label{MSfreq}
\end{equation}
where $\omega_{0}=\frac{\omega_{0}^1+\omega_{0}^2}{2}$ is the average clock transition carrier frequency, $\nu$ is the axial trap frequency, $\varepsilon$ is a symmetric detuning, and $\delta$ is an asymmetric detuning from the sideband transitions (see Fig. 2c,d). We also define the center laser frequency as $\omega_{L}=\frac{\omega_{+}+\omega_{-}}{2}=\omega_{0}-\delta$. Here, we work in the regime $\varepsilon\gg\eta\tilde{\Omega}$, where $\eta$ is the Lamb-Dicke parameter of the trap axial center-of-mass mode and $\tilde{\Omega}$ is the clock-transition  carrier Rabi frequency. In this regime, the coupling to motion through the red and blue sidebands can be adiabatically eliminated. In this case, two-photon coupling yields collective spin rotations and the Hamiltonian is well approximated as an Ising $\sigma_{y}\otimes\sigma_{y}$ interaction, with a $z$ transverse field due to $\delta_{1}=\delta$. If $\omega_{0}^1\ne\omega_{0}^2$, then dynamics is governed by the Hamiltonian in eq. \ref{Ising}, where $\delta_{2}$ represents the difference in detuning between ions and $\Omega=\frac{\eta^{2}\tilde{\Omega}^{2}}{\varepsilon}$ is the two-spin coupling. using the notation above, $\delta_1=\omega_{L}-\omega_{0}$ and $\delta_2=\frac{\omega_{0}^1-\omega_{0}^2}{2}$. Figure 2e,f shows a diagramatic illustration of the different detunings in this regime.\\

\begin{figure}[H]
\begin{subfigure}[t]{0.45\textwidth}
\includegraphics[width=\textwidth]{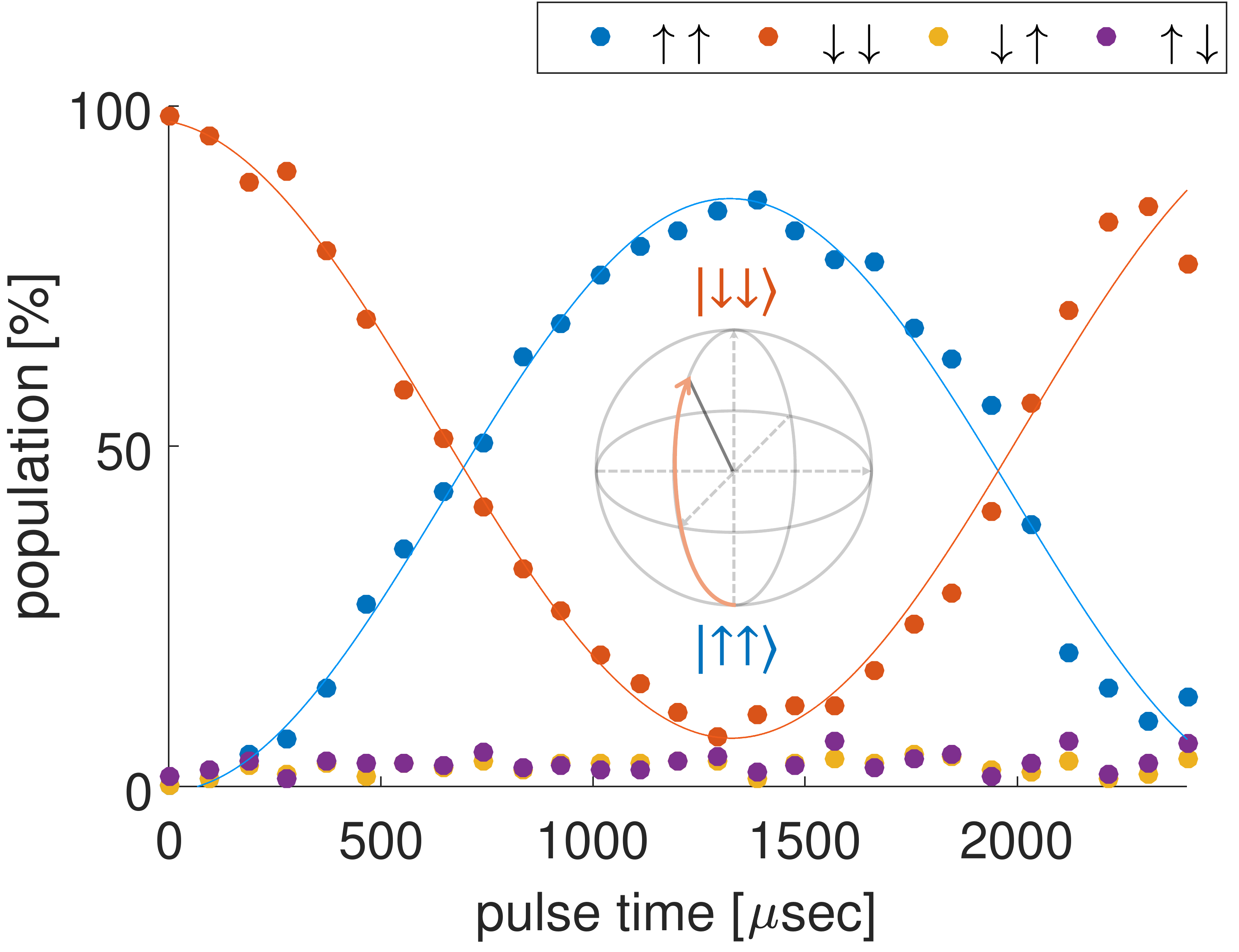}
\caption{}
\end{subfigure}
\begin{subfigure}[t]{0.45\textwidth}
\centering
\includegraphics[width=\textwidth]{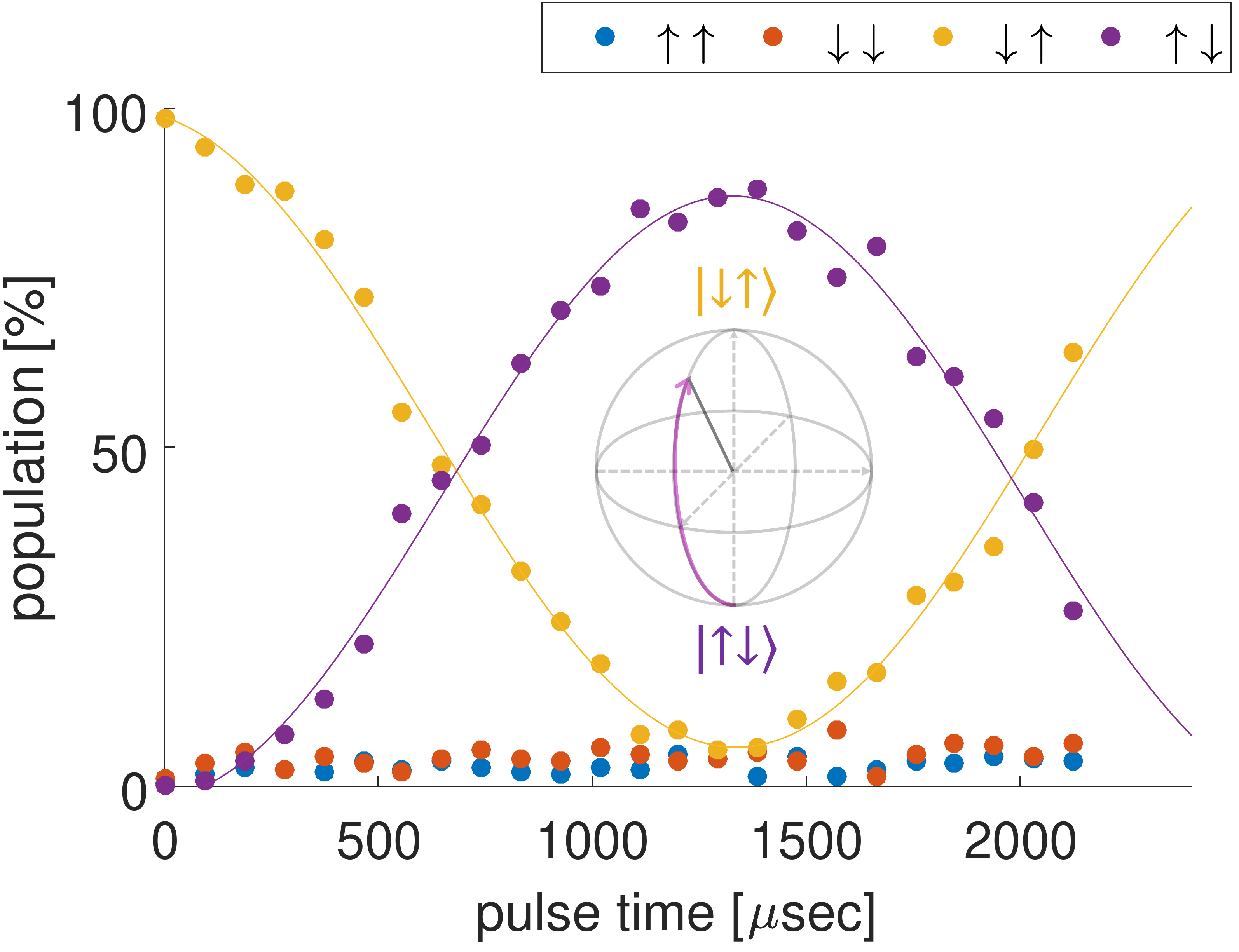}
\caption{}
\end{subfigure}
\hfill
\begin{subfigure}[t]{0.45\textwidth}
\centering
\includegraphics[width=\textwidth]{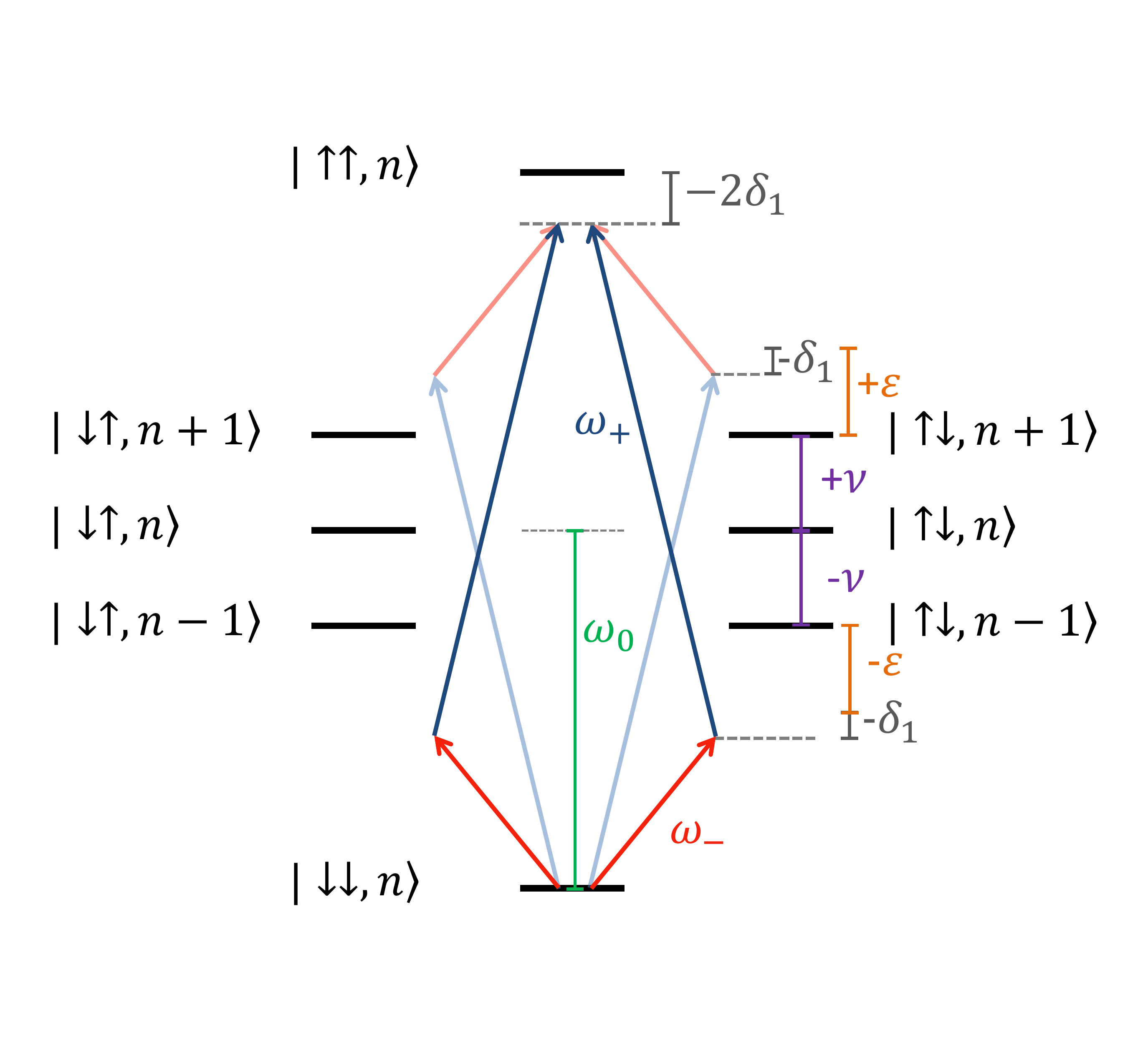}
\caption{}
\end{subfigure}
\begin{subfigure}[t]{0.45\textwidth}
\centering
\includegraphics[width=\textwidth]{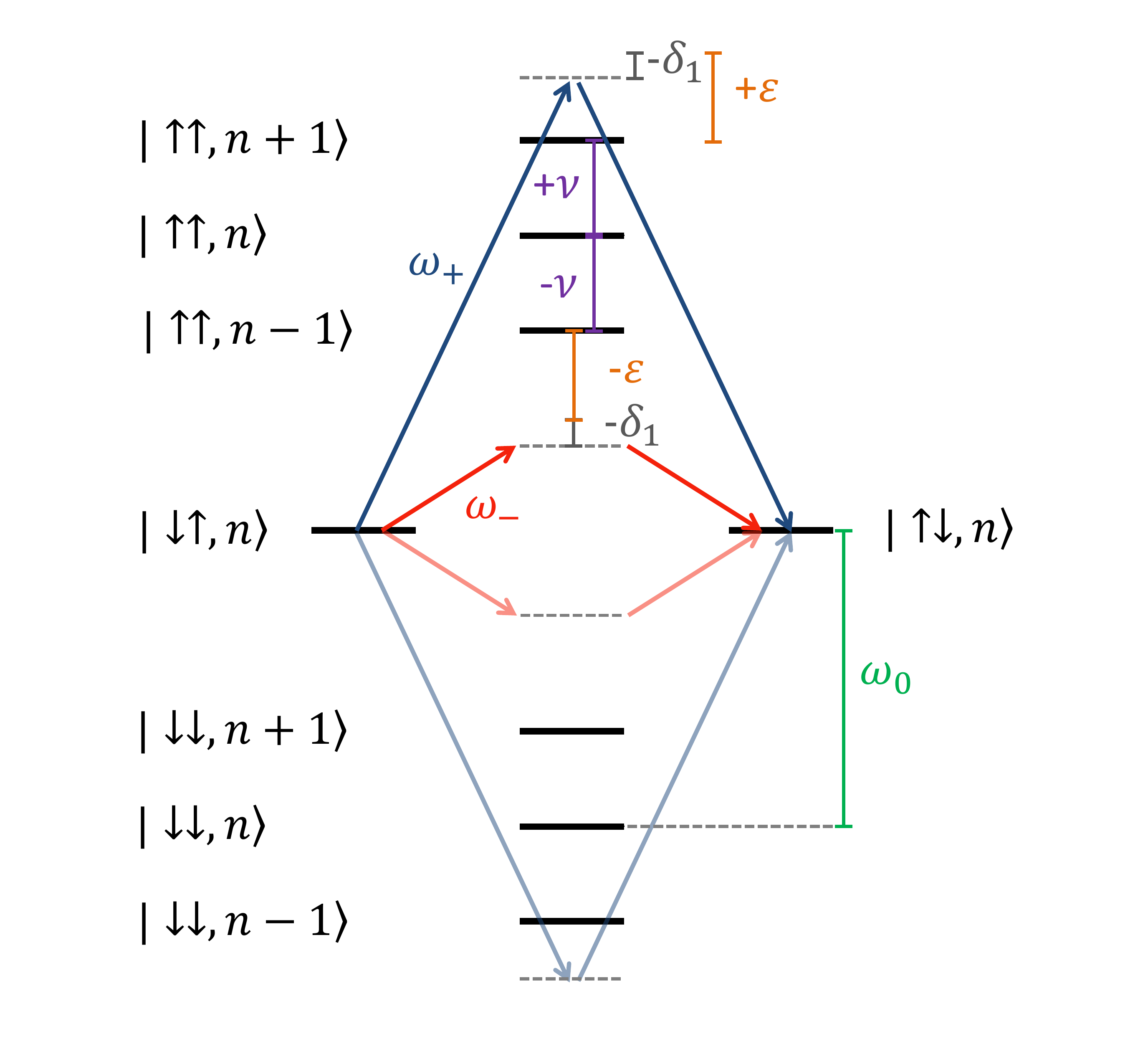}
\caption{}
\end{subfigure}
\hfill
\begin{subfigure}[t]{0.45\textwidth}
\centering
\includegraphics[width=\textwidth]{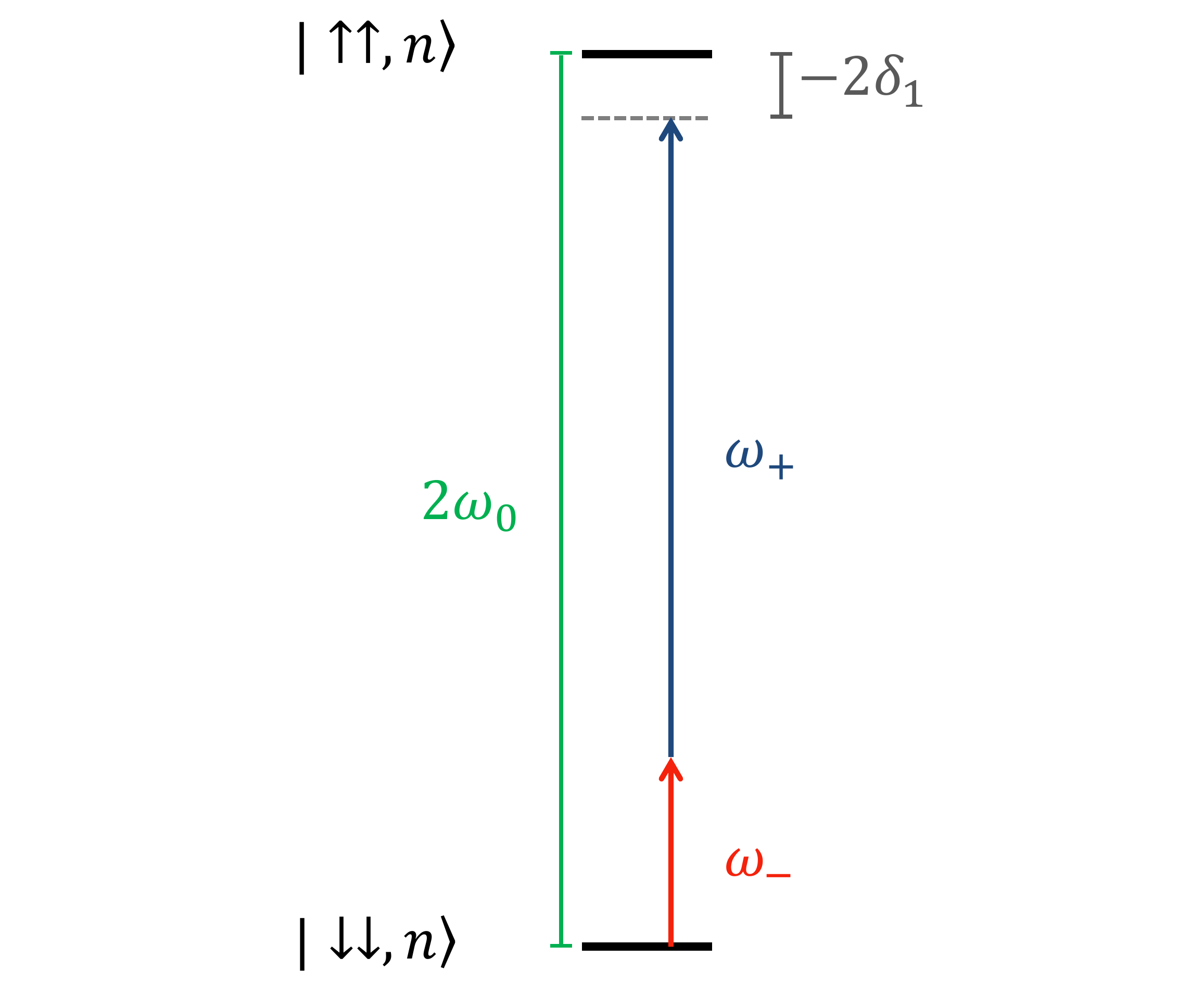}
\caption{}
\end{subfigure}
\hfill
\begin{subfigure}[t]{0.45\textwidth}
\centering
\includegraphics[width=\textwidth]{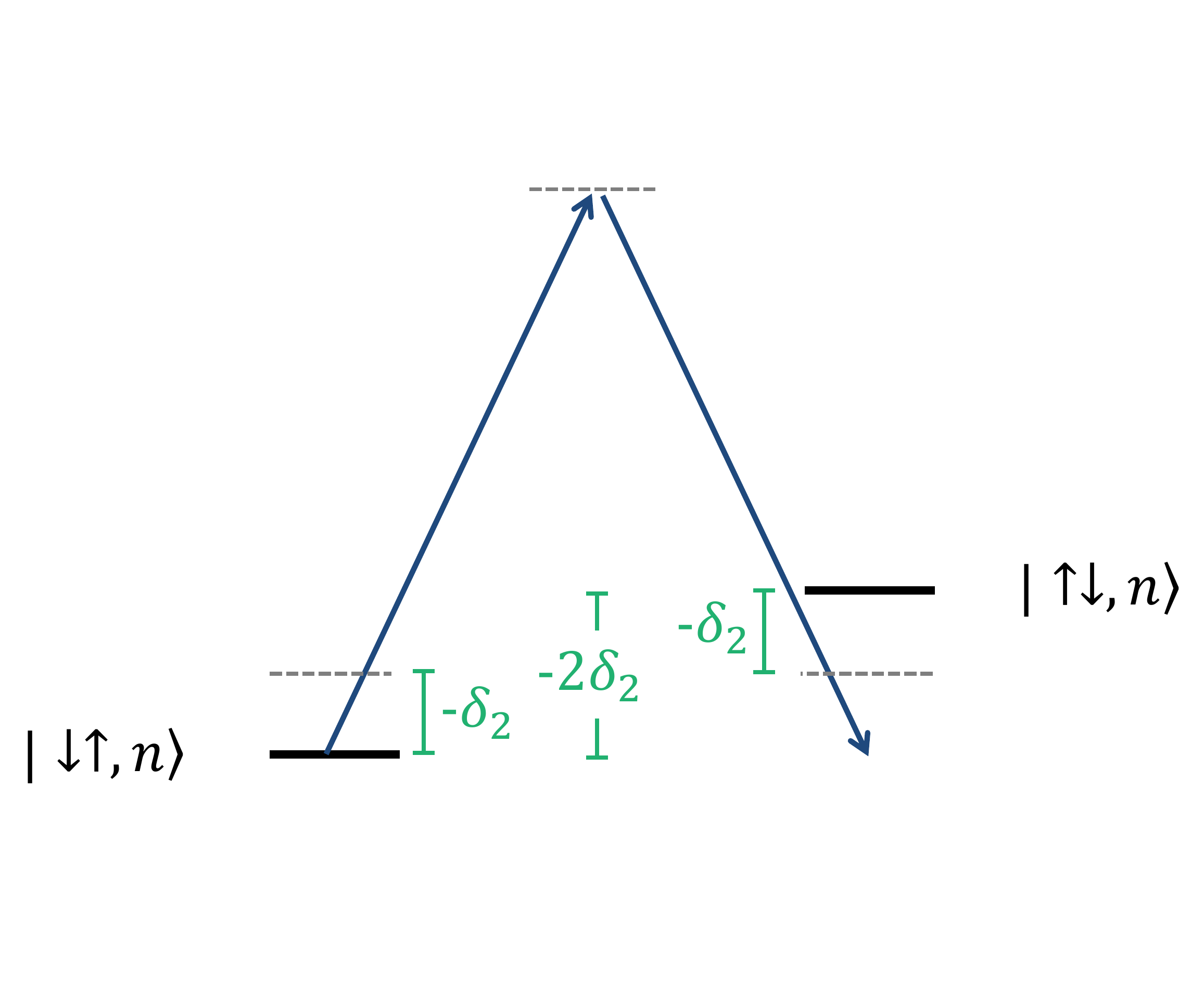}
\caption{}
\end{subfigure}
\caption{\textbf{Coupling in the two metrological subspaces using M\o lmer-S\o rensen interaction.} \textbf{(a,b)} Experimental results of resonant Rabi nutation between $\left|\downarrow\downarrow\right\rangle \leftrightarrow\left|\uparrow\uparrow\right\rangle $ and $\left|\uparrow\downarrow\right\rangle \leftrightarrow\left|\downarrow\uparrow\right\rangle $ respectively. Inset: Rotations illustrated on a Bloch sphere representing each subspace \textbf{(c,d)} Laser frequencies and the configuration  of energy levels coupled by the M\o lmer-S\o rensen operation for both $\left|\downarrow\downarrow\right\rangle \leftrightarrow\left|\uparrow\uparrow\right\rangle $ and $\left|\uparrow\downarrow\right\rangle \leftrightarrow\left|\downarrow\uparrow\right\rangle $ transitions respectively. In the limit of large $\varepsilon$ this operation approximates the Hamiltonian $H$ in Eq. \ref{Ising}. \textbf{(e,f)} A diagramatic representation of $\delta_{1}$ and $\delta_{2}$ scans in the two subspaces.}
\end{figure}

%%%%%%%%%%%%%%%%%%%%%%%%%%%%%%%%%%%%%%%%%%%%%%%%%%%%%%%%%%%%%%%%%
% Broad scan of delta_{1}
%%%%%%%%%%%%%%%%%%%%%%%%%%%%%%%%%%%%%%%%%%%%%%%%%%%%%%%%%%%%%%%%%

We begin by performing correlated Rabi nutation in the two subspaces using resonant Ising interaction. Here, we initialized our system in $\left|\downarrow\uparrow\right\rangle $ or $\left|\downarrow\downarrow\right\rangle $ and turned-on our MS interaction setting $\delta_{1} \simeq \delta_{2} \simeq 0$. Correlated Rabi nutation curves in the two subspaces are shown in Fig. 2a,b. As seen, coupling to states outside the subspace is minimized by the choice of large $\varepsilon$. We observe a complete correlated spin-flip at a $\pi$-time of $\tau_{\pi}=2\frac{2\pi\varepsilon}{\eta^{2}\Omega^{2}}$, which is about $1300\ \mu sec$ in this experiment.

Next we performed a wide Rabi spectroscopy scan by scanning $\delta_{1}$ from $-\varepsilon$ to $\varepsilon$; i.e. nearly to the motional sideband; by scanning the MS laser center frequency, $\omega_L$. A measurement of the populations of all four spin states vs. $\delta_{1}$ is shown in Fig. 3a-d. Here we set $\delta_{2}\simeq 0$ and the pulse time to $\tau_{\pi}$. As seen, when the system is initialized in the even subspace correlated spin rotation does not occur unless $\delta_{1}  \simeq 0$. Around this resonant value, marked by a grey background, a sharp $\left|\downarrow\downarrow\right\rangle\rightarrow\left|\uparrow\uparrow\right\rangle$ transition is observed. This correlated spin-flip resonance is enlarged in the inset of Figures 3c,d. On the other hand, when the system is initialized in the odd subspace, correlated spin-flip occurs at any value of $\delta_{1}$. This is due to the fact that the $\{\left|\downarrow\uparrow\right\rangle\,\left|\uparrow\downarrow\right\rangle\}$ subspace is insensitive to $\delta_{1}$. This odd subspace has been used several times before as a decoherence-free subspace due to this resilience to common phase noise. In both subspaces, as $\delta_{1}$ approaches $\varepsilon$, single-photon sideband transitions occur resulting in rapid population oscillations. \\
%%%%%%%%%%%%%%%%%%%%%%%%%%%%%%%%%%%%%%%%%%%%%%%%%%%%%%%%%%%%%%%%%
% light shift scan
%%%%%%%%%%%%%%%%%%%%%%%%%%%%%%%%%%%%%%%%%%%%%%%%%%%%%%%%%%%%%%%%%
\begin{figure}[H]
\begin{subfigure}[t]{0.45\textwidth}
\includegraphics[width=\textwidth]{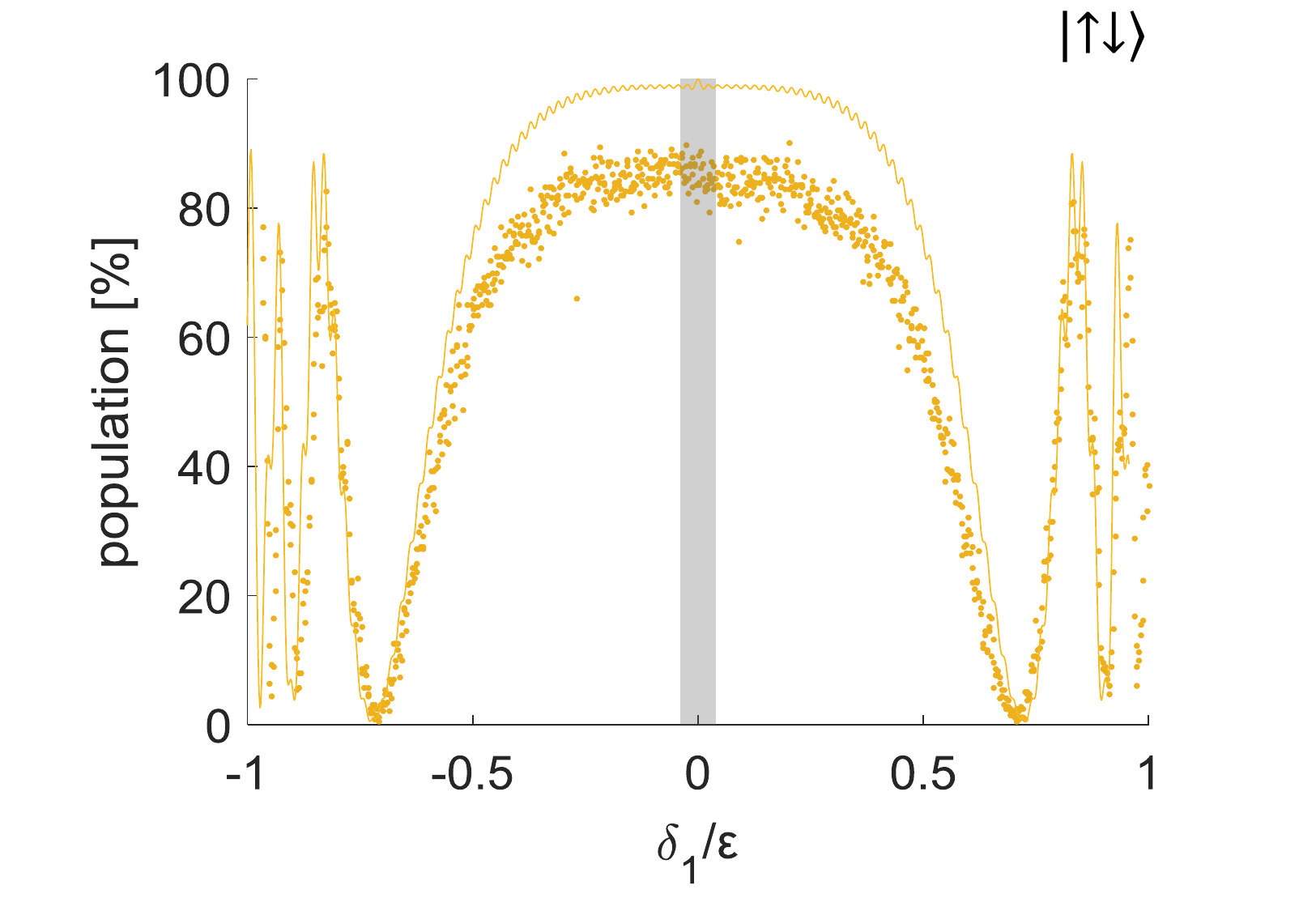}
\caption{}
\end{subfigure}
\begin{subfigure}[t]{0.45\textwidth}
\centering
\includegraphics[width=\textwidth]{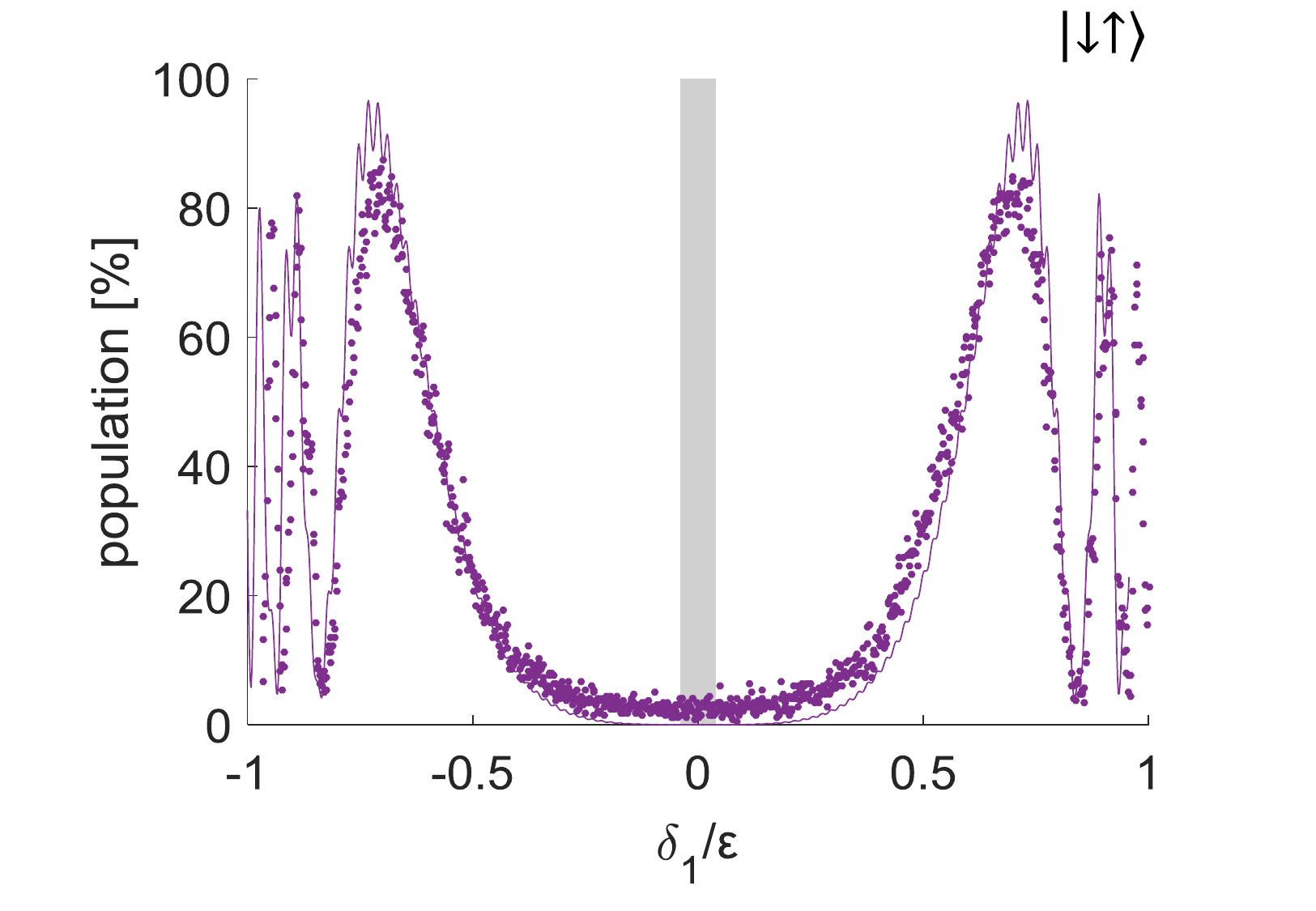}
\caption{}
\end{subfigure}
\hfill
\begin{subfigure}[t]{0.45\textwidth}
\centering
\includegraphics[width=\textwidth]{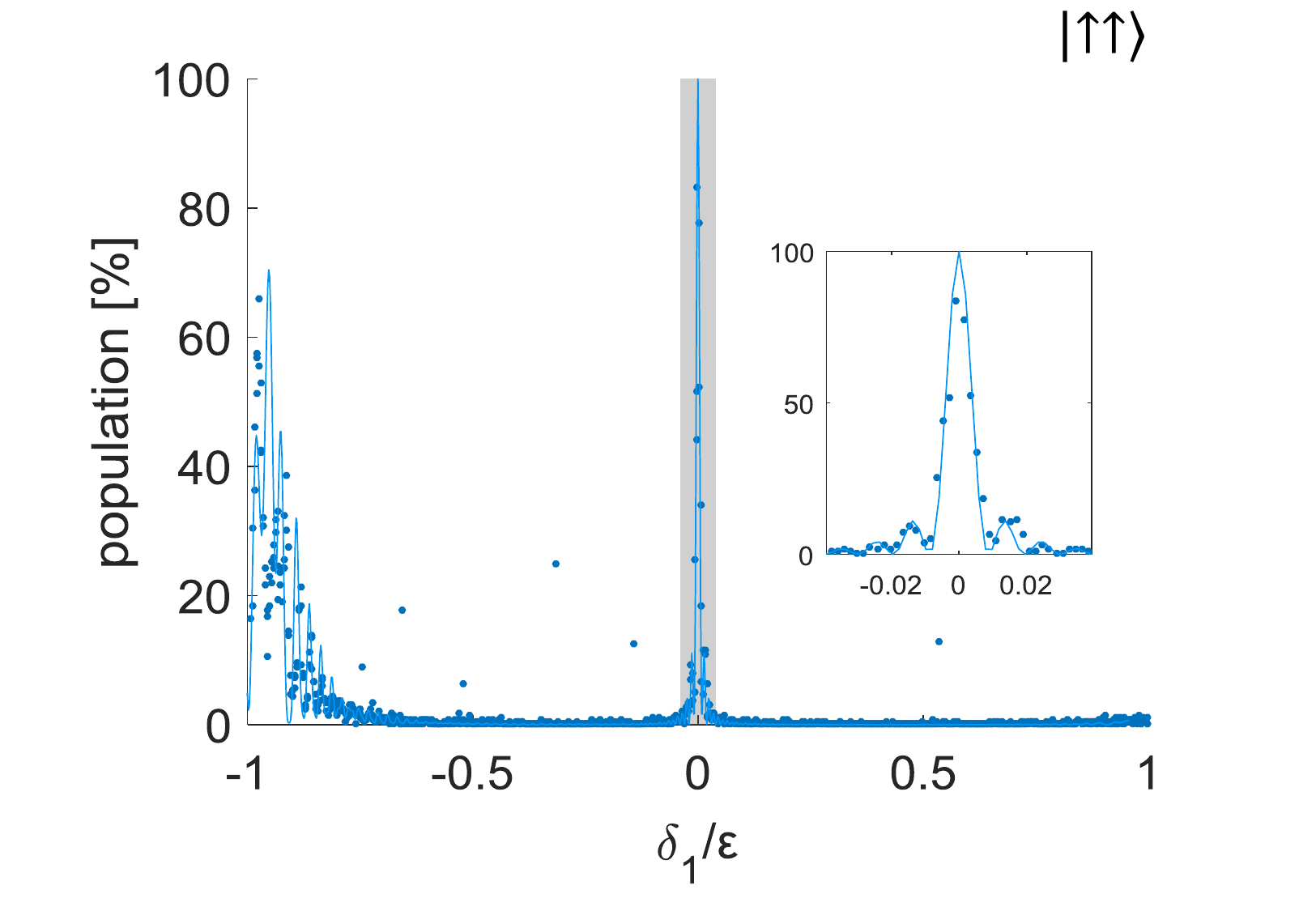}
\caption{}
\end{subfigure}
\hfill
\begin{subfigure}[t]{0.45\textwidth}
\centering
\includegraphics[width=\textwidth]{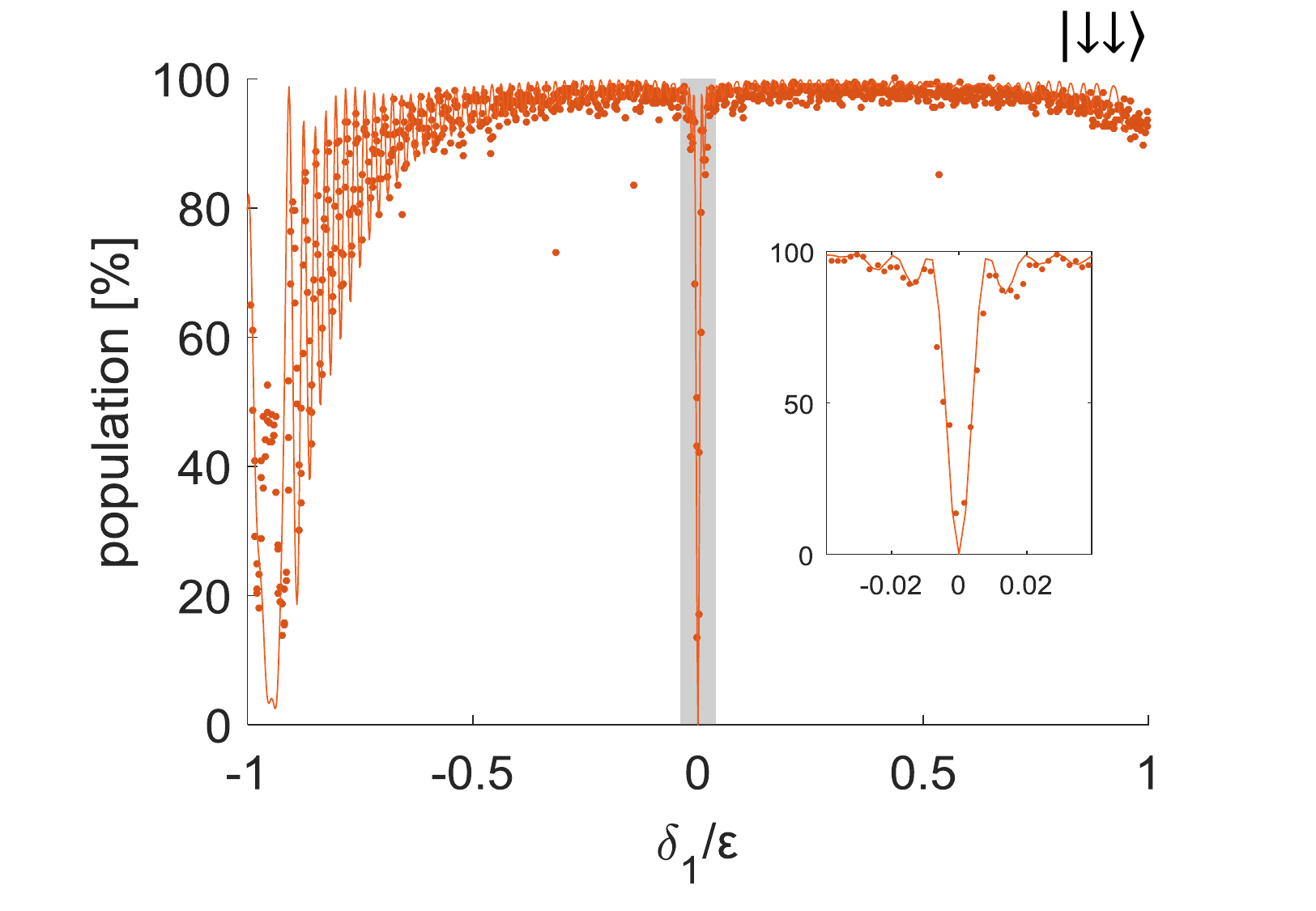}
\caption{}
\end{subfigure}
\caption{\textbf{Broad scan of $\delta_{1}$ in both subspaces} \textbf{(a,b)} $\delta_{1}$ scan results when initializing in $\left|\downarrow\uparrow\right\rangle$ and measuring the population in $\left|\uparrow\downarrow\right\rangle$ and $\left|\downarrow\uparrow\right\rangle$ respectively. \textbf{(c,d)} $\delta_{1}$ scan results when initializing in $\left|\downarrow\downarrow\right\rangle$ and measuring the population in $\left|\uparrow\uparrow\right\rangle$ and $\left|\downarrow\downarrow\right\rangle$ respectively. The data was shifted to be symmetric around $\delta_{1}=0$, and the solid lines are a simulation results with no fit parameters. The discrepancy between simulation and data in (a) is due to in population leaking to ${\left|\downarrow\downarrow\right\rangle,\left|\uparrow\uparrow\right\rangle}$ owing to experimental imperfections. The shaded grey area illustrates the difference in sensitivity to $\delta_{1}$ in both subspaces, and the insets of (c) and (d) are magnification of that scan interval. In both measurements $\varepsilon = 2\pi \times 25.5 \mathrm{kHz} \approx 10\eta\Omega$.}
\end{figure}

We next turned to a combined scan of both $\delta_{1}$ and $\delta_{2}$. This scan was created by light-shifting the resonance frequency of only one of the ions by using an off-resonance single-addressing beam (see Fig. 2). With a detuning $\delta_{ls}/2\pi \simeq 3.5 \mathrm{MHz}$ and a Rabi-frequency which varied between $\Omega_{ls}/2\pi \simeq 0 - 40 \mathrm{kHz}$ we scanned the light-shift between $\Delta f_{ls}/2\pi\approx\frac{\Omega_{ls}^{2}}{2\pi\delta_{ls}}\simeq 0 - 400 \mathrm{Hz}$.

\begin{figure}[H]
\begin{subfigure}[t]{0.35\textwidth}
\includegraphics[width=\textwidth]{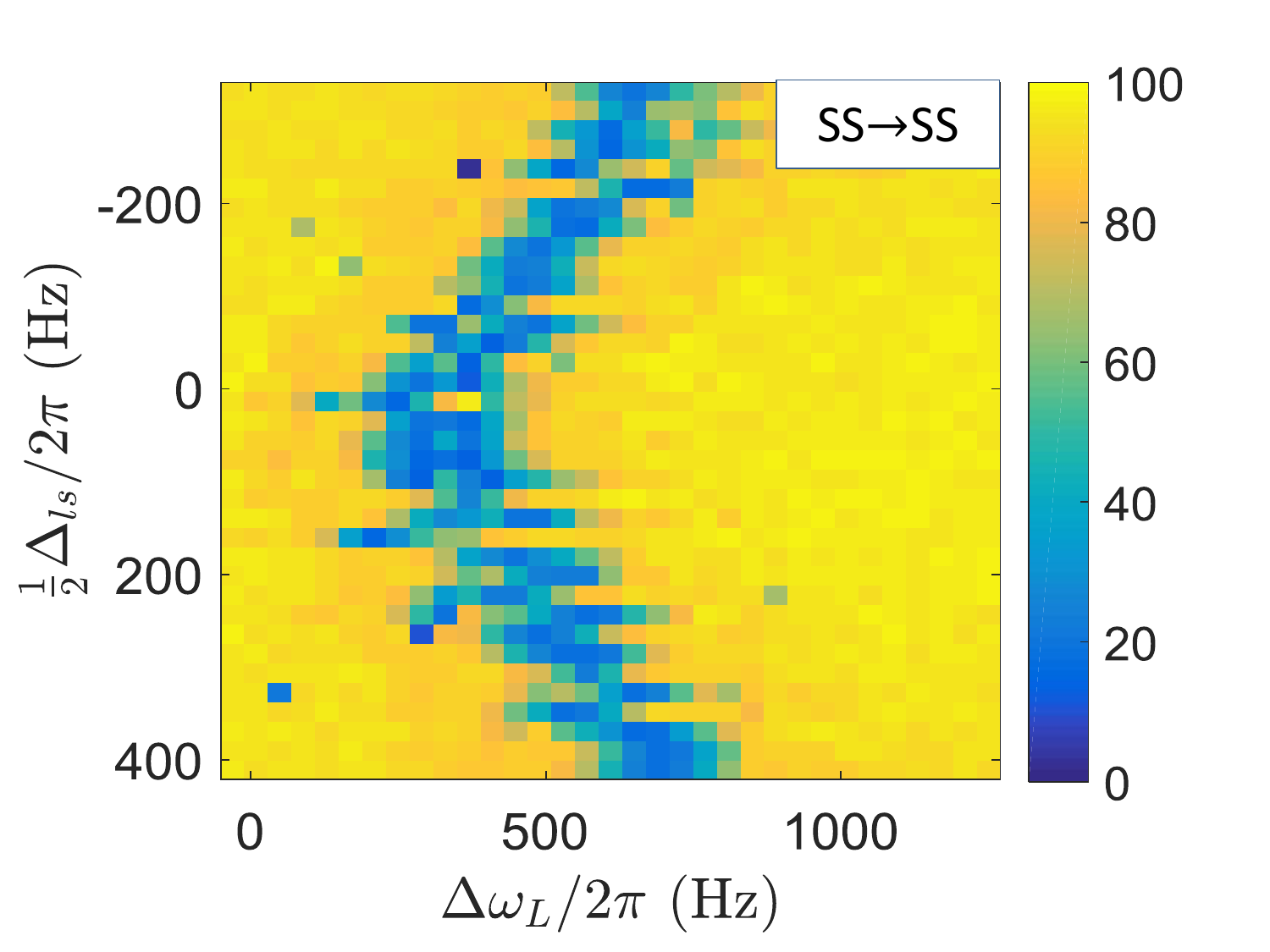}
\caption{}
\end{subfigure}
\begin{subfigure}[t]{0.35\textwidth}
\centering
\includegraphics[width=\textwidth]{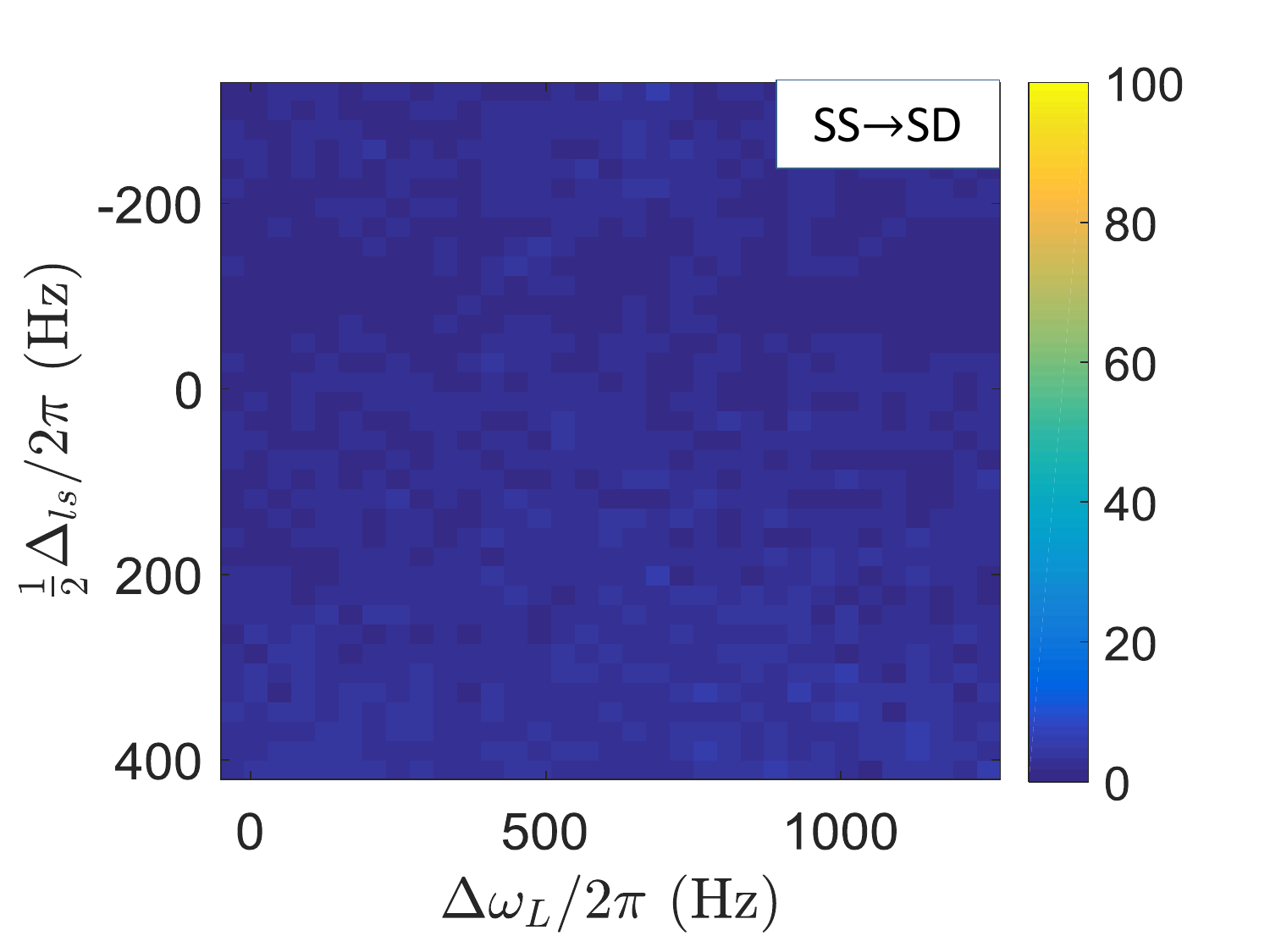}
\caption{}
\end{subfigure}
%\hfill
\begin{subfigure}[t]{0.35\textwidth}
\centering
\includegraphics[width=\textwidth]{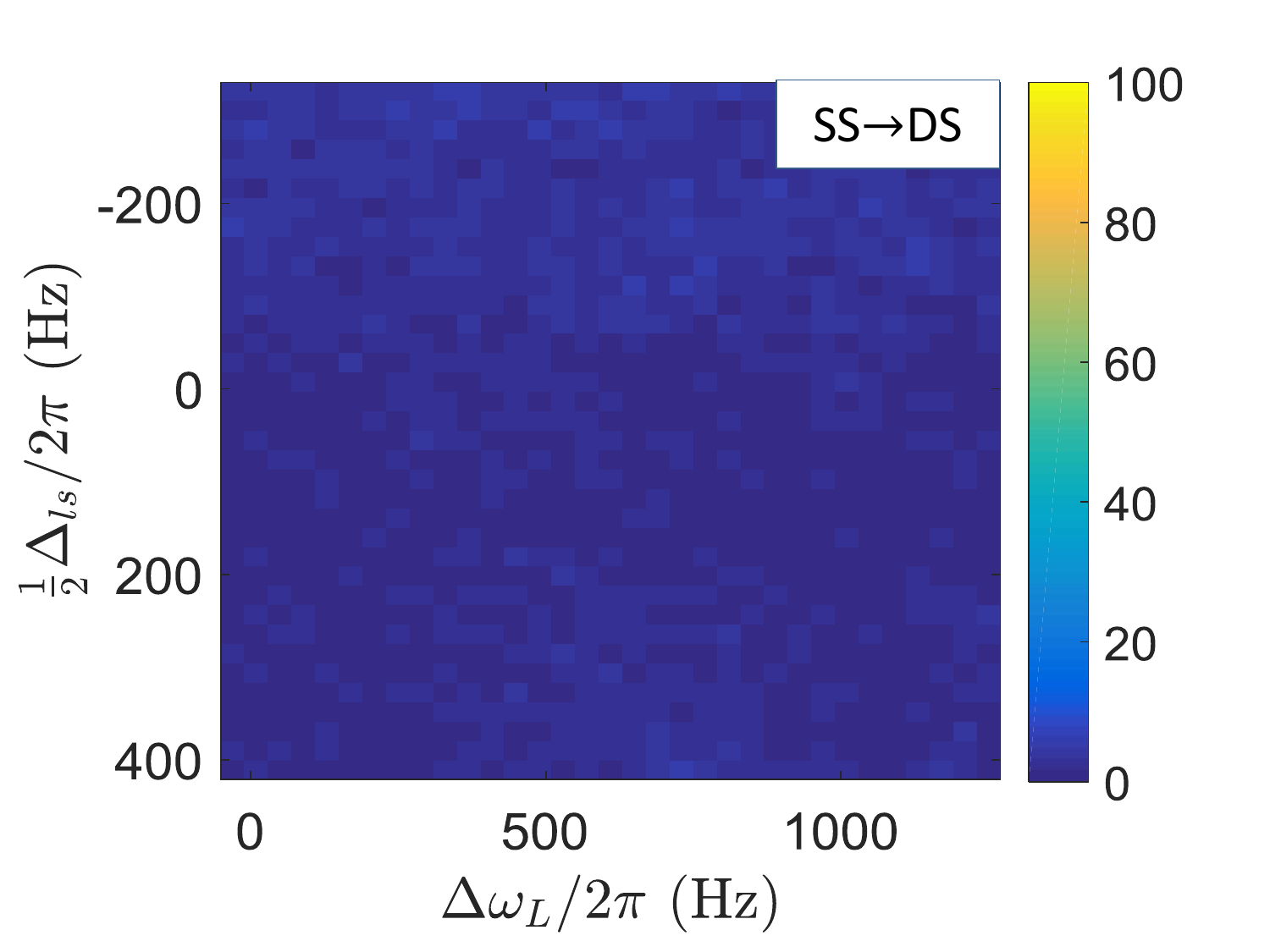}
\caption{}
\end{subfigure}
%\hfill
\begin{subfigure}[t]{0.35\textwidth}
\centering
\includegraphics[width=\textwidth]{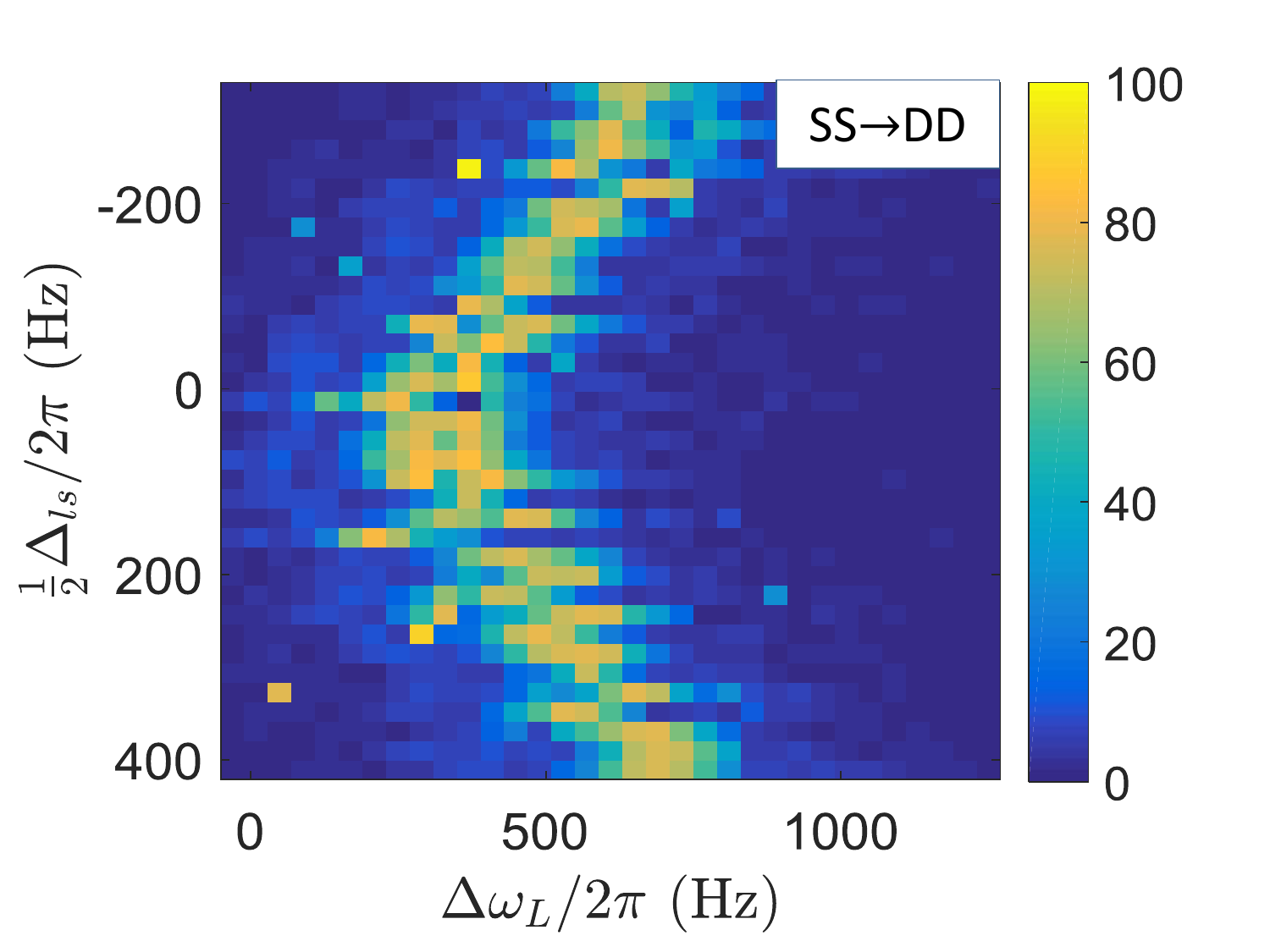}
\caption{}
\end{subfigure}
%\hfill
\begin{subfigure}[t]{0.35\textwidth}
\centering
\includegraphics[width=\textwidth]{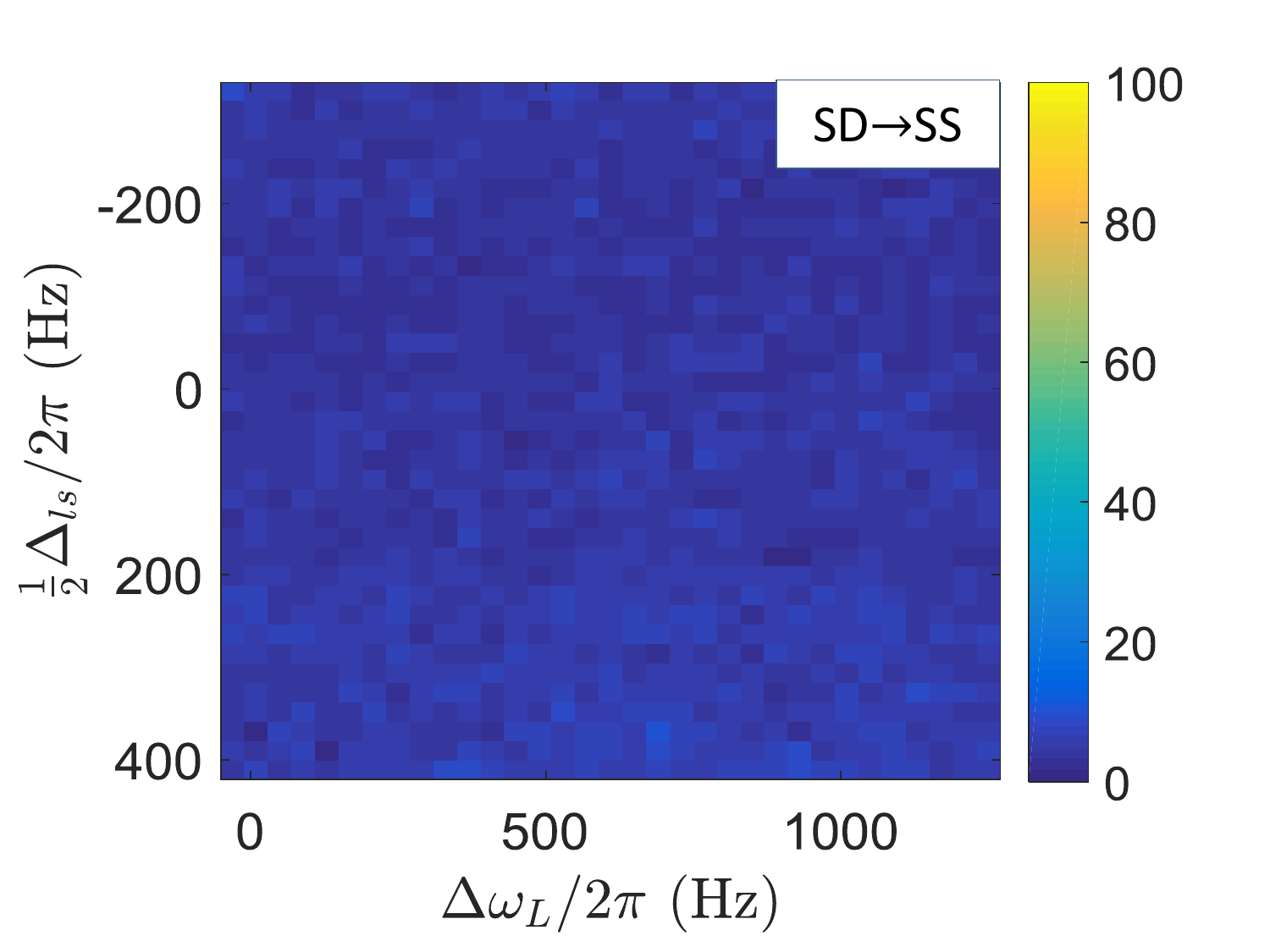}
\caption{}
\end{subfigure}
%\hfill
\begin{subfigure}[t]{0.35\textwidth}
\centering
\includegraphics[width=\textwidth]{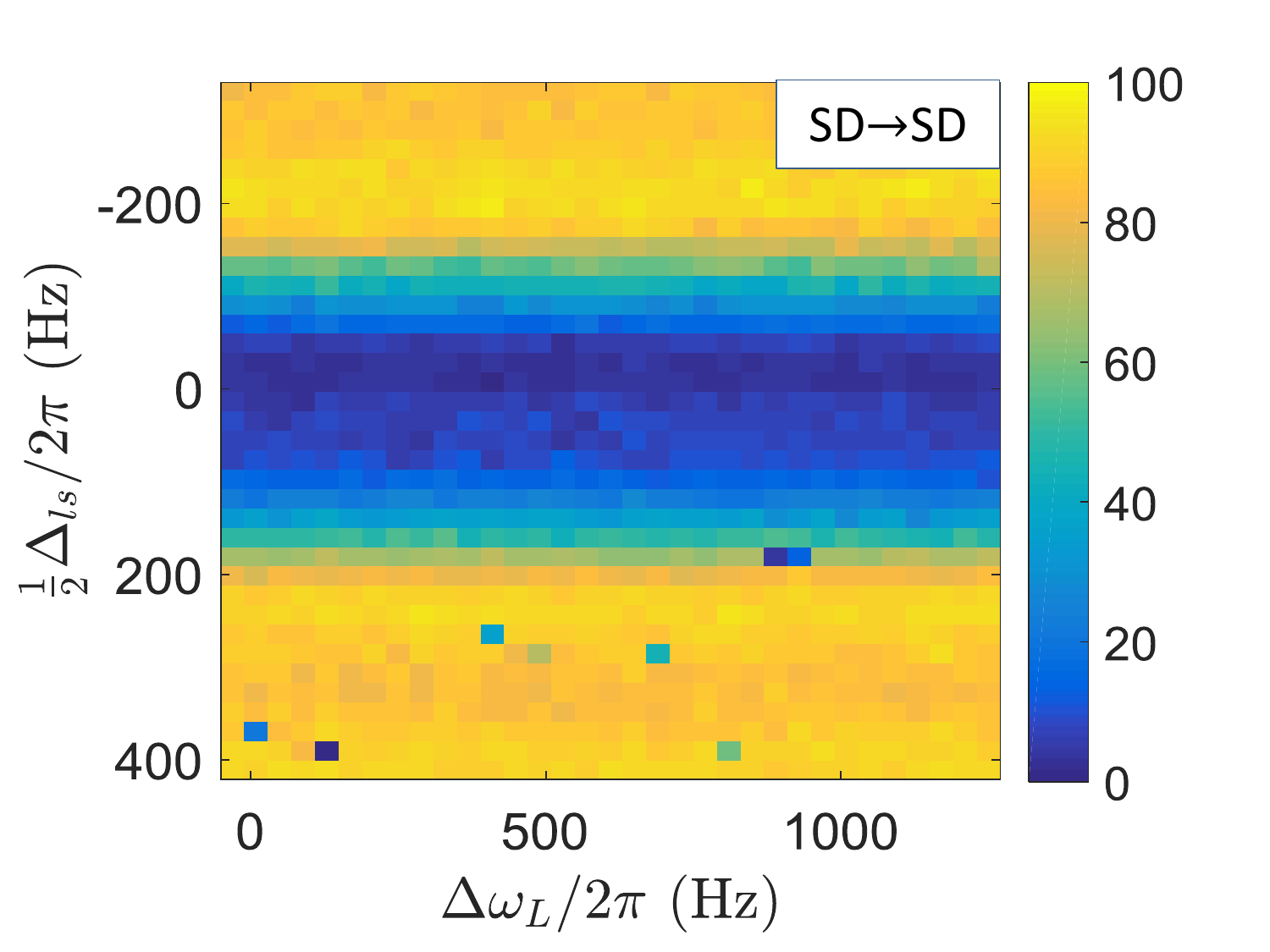}
\caption{}
\end{subfigure}
%\hfill
\begin{subfigure}[t]{0.35\textwidth}
\centering
\includegraphics[width=\textwidth]{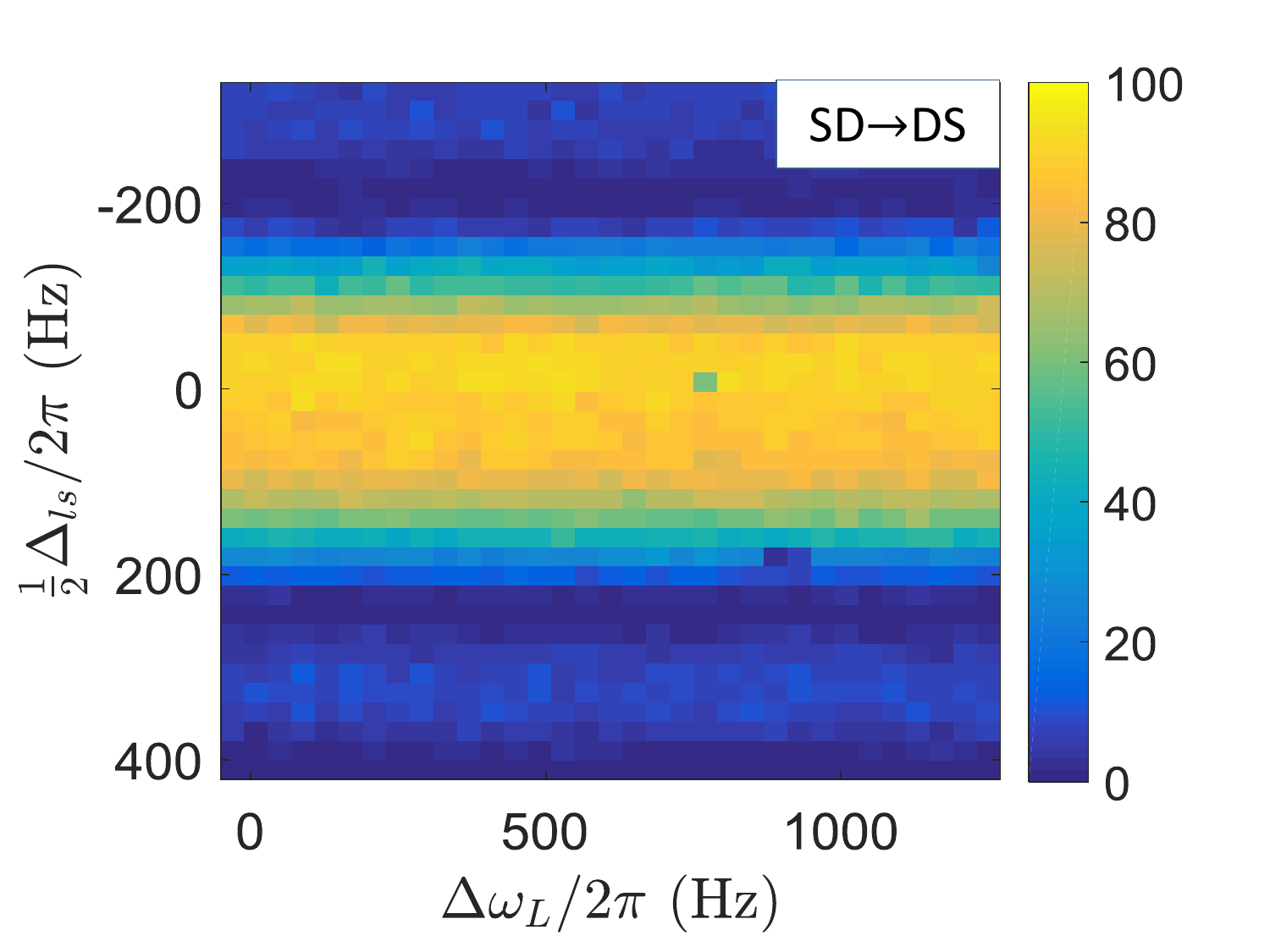}
\caption{}
\end{subfigure}
\hfill
\begin{subfigure}[t]{0.35\textwidth}
\centering
\includegraphics[width=\textwidth]{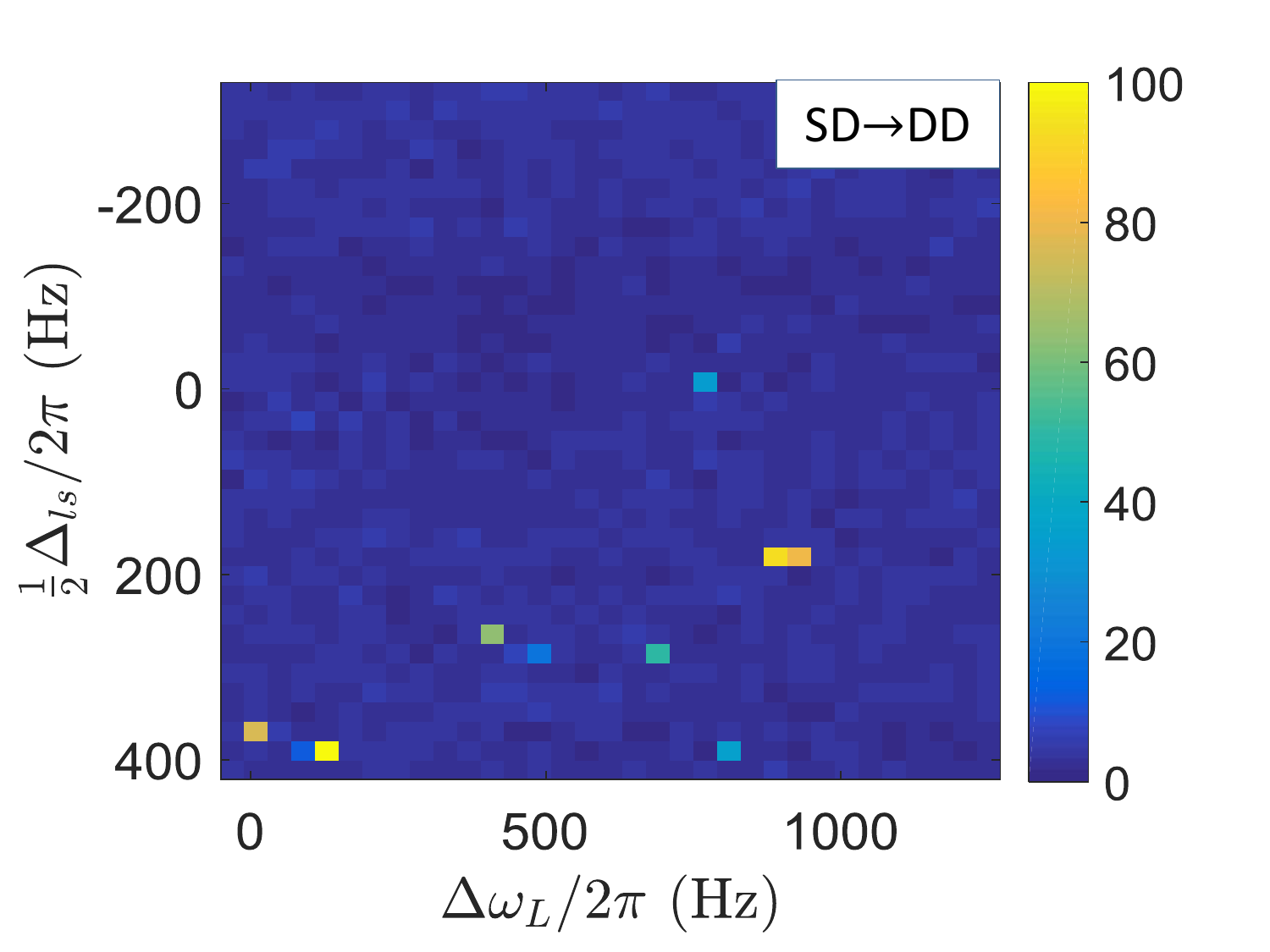}
\caption{}
\end{subfigure}
\caption{\textbf{Sensitivity of correlated rotations to $\delta_{1}$ and $\delta_{2}$}. \textbf{(a,b,c,d)} Population of $\left|\downarrow\downarrow\right\rangle$, $\left|\downarrow\uparrow\right\rangle$, $\left|\uparrow\downarrow\right\rangle$, $\left|\uparrow\uparrow\right\rangle$ when initializing in $\left|\downarrow\downarrow\right\rangle$ respectively. \textbf{(e,f,g,h)} Population of $\left|\downarrow\downarrow\right\rangle$, $\left|\downarrow\uparrow\right\rangle$, $\left|\uparrow\downarrow\right\rangle$, $\left|\uparrow\uparrow\right\rangle$ when initializing in $\left|\downarrow\uparrow\right\rangle$ respectively. We shift the resonance frequency of one of the two ions by $\Delta f_{ls}$ using a tightly focused, off-resonance, laser beam. As a result, $\delta_{1}$ is shifted by $\left|\frac{\Delta f_{ls}}{2}\right|$ and $\delta_{2}$ is shifted by $\pm\frac{\Delta f_{ls}}{2}$, where the sign depends which ion is shifted. For each value of  $\Delta f_{ls}$ we scan the laser frequency $\omega_L$ with respect to an arbitrary offset and measure populations. As shown a resonance in the odd subspace appears every time $\delta_{2}=0$ regardless of $\omega_L$ which only shifts $\delta_{1}$. On the other hand in the even subspace $\Delta f_{ls}$ shifts the $\delta_{1}$ resonance symmetrically when either of the ions is light shifted, indicating that the associated change in  $\delta_{2}$ does not affect this subspace.}
\end{figure}

By definition, $\delta_{2}=\pm\frac{1}{2}\Delta_{ls}/2\pi$. The sign is determined by the specific ion being light-shifted. The magnitude of $\Delta f_{ls}$, and therefore $\delta_{2}$, was scanned by varying the intensity of the individual addressing laser. For every value of $\Delta f_{ls}$ a full scan of $\omega_L$ was carried out, by changing the parameter $\delta$ in eq. \ref{MSfreq}. Figure 4 shows the population of all spin states for such a scan, when initializing in $\left|\downarrow\downarrow\right\rangle$ (a-d) and in $\left|\downarrow\uparrow\right\rangle$ (e-h). As seen, in the odd subspace, a change to $\Delta f_{ls}$ causes a resonant response every time $\delta_2=0$, whereas a change of $\omega_L$ does not change the position of this resonance. In the even subspace a scan of $\omega_L$ yields a resonant response every time $\omega_L-\omega_{0}=0$. The change of $\Delta f_{ls}$ shifts the position of this resonance symmetrically with respect to the sign of $\Delta f_{ls}$, leading to the curved shape is figures 4a and 4d. This symmetry proves that in the symmetric subspace it is only the contribution of $\Delta f_{ls}$ to $\delta_1$ which changes the resonance position (see supplementary material). Note that since symmetric phase noise is more common in our experiment than differential phase noise between the two ions, the resonance in the symmetric sub-space is much noisier than in the anti-symmetric subspace.

\begin{figure}[H]
\begin{subfigure}[t]{0.5\textwidth}
\includegraphics[width=\textwidth]{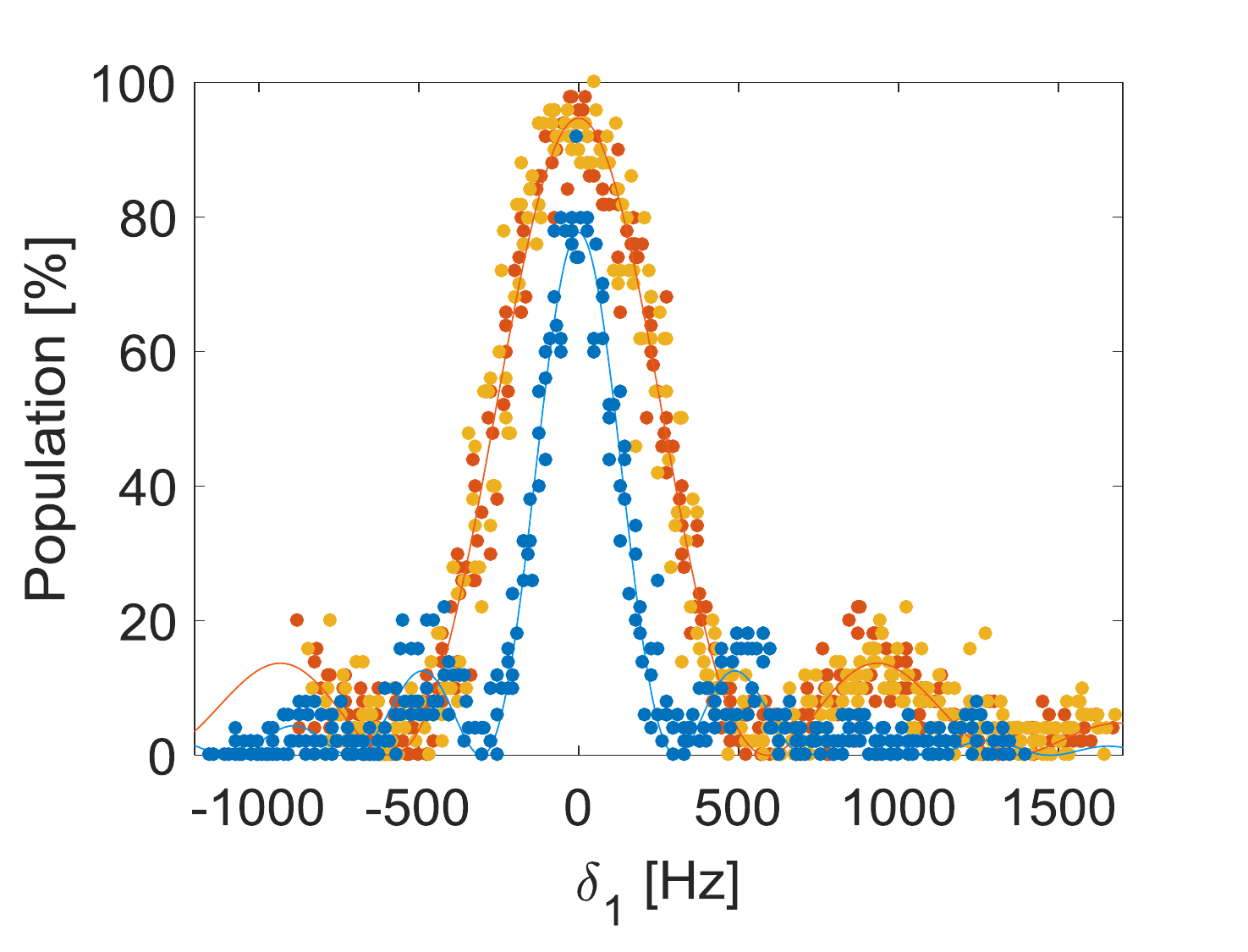}
\caption{}
\end{subfigure}
\hfill
\begin{subfigure}[t]{0.5\textwidth}
\includegraphics[width=\textwidth]{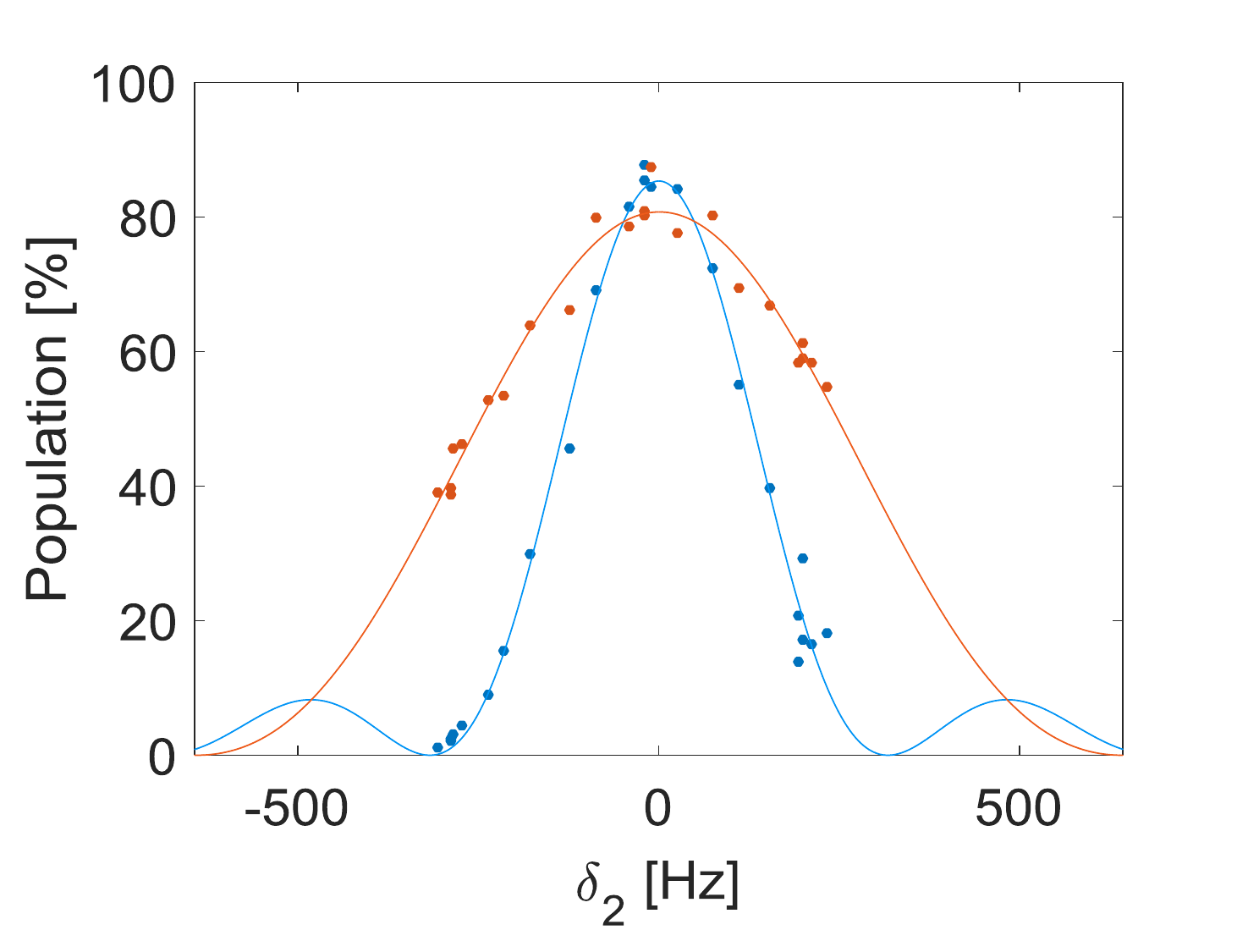}
\caption{}
\end{subfigure}
\hfill
\begin{subfigure}[t]{0.5\textwidth}
\includegraphics[width=\textwidth]{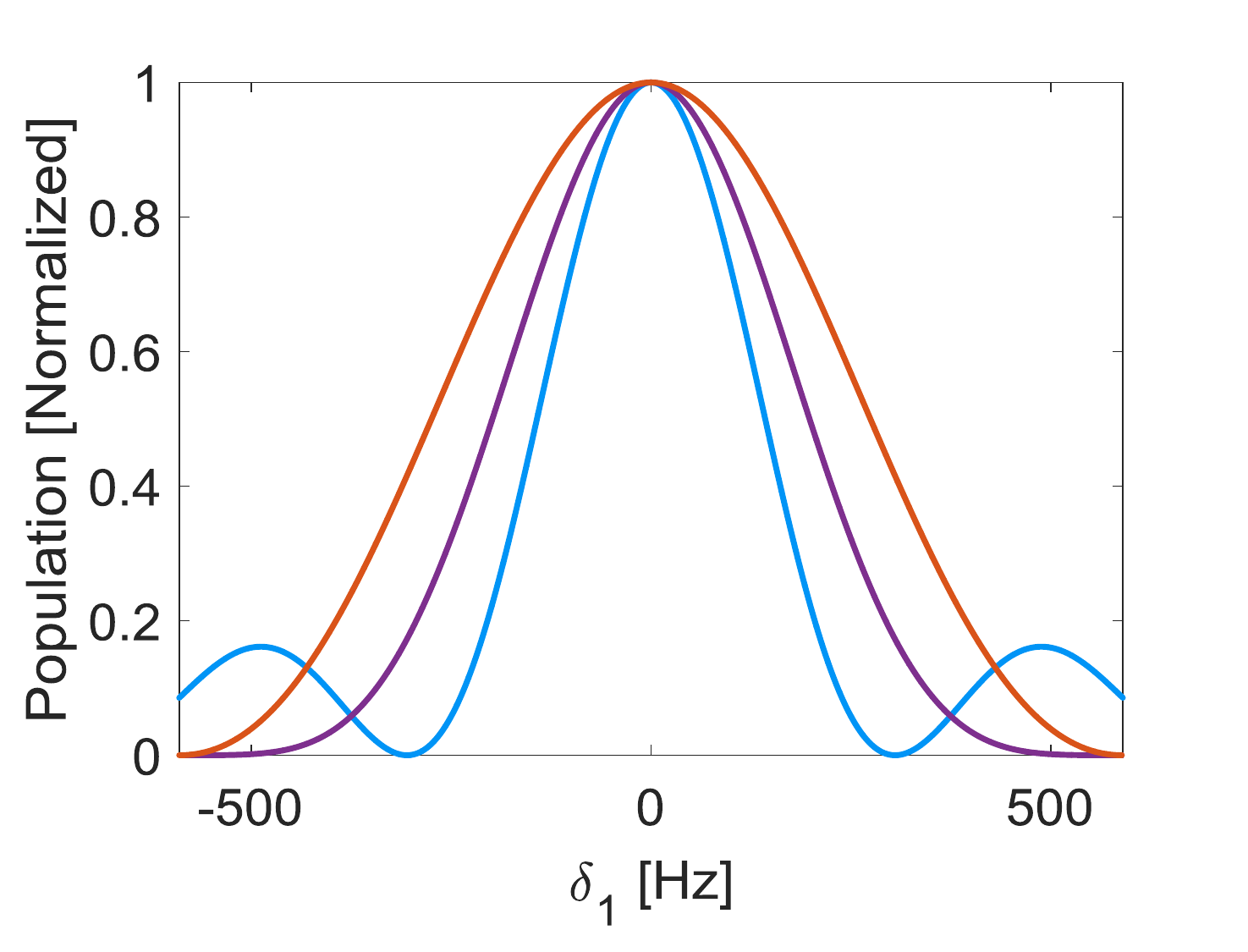}
\caption{}
\end{subfigure}
\hfill
\begin{subfigure}[t]{0.5\textwidth}
\includegraphics[width=\textwidth]{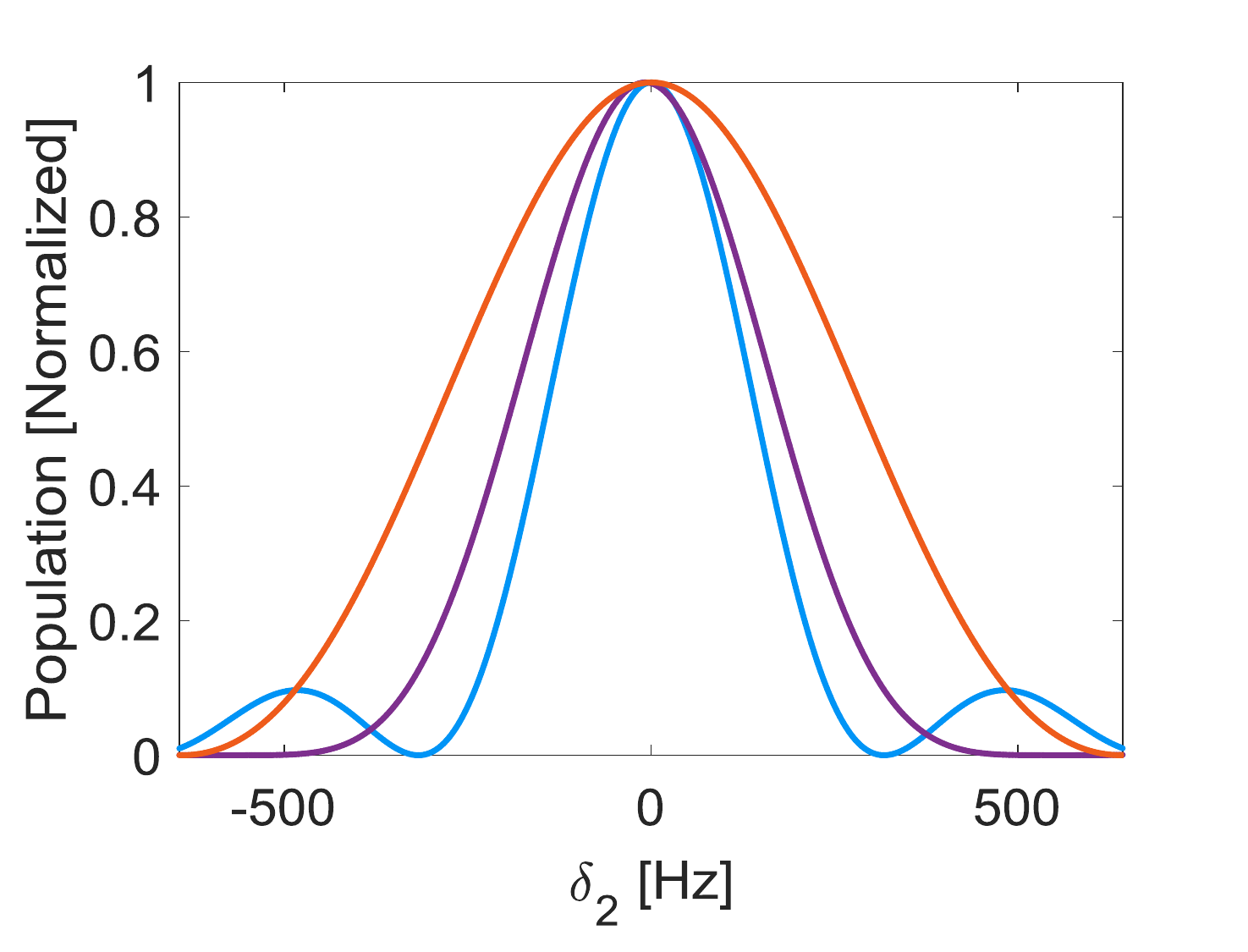}
\caption{}
\end{subfigure}
\caption{Heisenberg-limited spectroscopy. \textbf{(a)} Comparison between the spectra for the cases of a single ion (red and orange full circles for the right and left ion respectively) and two ions correlated spectroscopy (blue full circles). All spectra were shifted to center around $\delta_{1}=0$ due to small shifts ($10's$ $\mathrm{Hz}$) and fitted to Eq.3 (solid blue line). \textbf{(b)} Comparison between the frequency-difference spectra of two ions in an uncorrelated (red full circles) and a correlated (blue full circles) measurement. In the uncorrelated case the average frequency of the two ions was measured on each ion separately as a function of $\delta_{2}$, and both populations were averaged to give the measurement result (See supplementary material). The spectra were also shifted to center around $\delta_{2}=0$. \textbf{(c,d)} Comparison between the normalized uncorrelated case fit, the product of the two single ion fits in the uncorrelated case and the normalized correlated fit for both $\delta_{1}$ and $\delta_{2}$ scans respectively. As seen, the single ion case is the widest, but holds two data point for each scanned parameter value, and therefore estimation of the resonance frequency can be performed with $\sqrt{2}$ reduction in the uncertainty. This can be manifested spectrum-wise by taking a spectrum of the probability of both ions excited (solid purple line). The spectrum is $\sqrt{2}$ narrower than the single ion case, but not all measurements are included, thus matching the resonance frequency uncertainty to the two, uncorrelated, ions case. The correlated case clearly yields the narrowest spectrum with the same number of significant measurements and therefore a $\sqrt{2}$ improvement in the frequency estimation uncertainty due to shot-noise.}
\end{figure}
Finally, to determine whether our spectroscopy is Heisenberg-limited we performed a narrower scan of $\delta_{1}$ and $\delta_{2}$, each in its corresponding subspace, and compared the resulting spectrum to that of standard Rabi spectroscopy. Here $\delta_{1}$ scan in the even subspace was performed by scanning $\omega_L$, and $\delta_{2}$ scan in the odd subspace was achieved by scanning $\Delta f_{ls}$. Uncorrelated Rabi spectroscopy was performed by a regular single-ion Rabi spectrocopy scan.

The spectral shape of the excited-state population in  two-level Rabi spectroscopy is \cite{AllenEberly1975},
\begin{equation}
\begin{split}
P(\uparrow)=A\frac{\sin^{2}\left(\frac{\Omega\tau}{2}\sqrt{1+\left(\frac{\alpha\delta}{\Omega}\right)^{2}}\right)}{1+\left(\frac{\alpha\delta}{\Omega}\right)^{2}}, \end{split}\label{spectrum}
\end{equation}
where $\Omega$ is the Rabi frequency and the contrast parameter $A$ accounts for experiment imperfections such as dephasing or noise in the Rabi coupling. $\alpha$ is the narrowing factor which is $1$ for a uncorrelated Rabi spectroscopy and $2$ for perfect two-qubit correlated Heisenberg limited Rabi spectroscopy.

The results of the different scans are shown in Fig. 5. For the $\delta_{1}$ scan in the even subspace we obtained $\alpha=1.92\pm0.02$ for a correlated rotation and $\alpha=1.01\pm0.01$ for the single ion case using a maximum likelihood fit to Eq.\ref{spectrum}. In the odd subspace, a $\delta_{2}$ scan yielded $\alpha=1.78\pm0.03$ for the correlated case and $f=0.88\pm0.02$ for single ion spectroscopy. These results show that using correlated rotation we are indeed well below the standard quantum limit and close to the Heisenberg limit of frequency estimation.\\

%%%%%%%%%%%%%%%%%%%%%%%%%%%%%%%%%%%%%%%%%%%%%%%%%%%%%%%%%%%%%%%%%
% Discussion
%%%%%%%%%%%%%%%%%%%%%%%%%%%%%%%%%%%%%%%%%%%%%%%%%%%%%%%%%%%%%%%%%
%As the Spectroscopy shown here exhibits narrower spectrum compared with classical multi-reference spectroscopy, it is inferred that using entanglement one can decrease the uncertainty in frequency measurements. This is indeed the case for a shot-noise limited measurement, as the uncertainty lowers with more averaging, and in the correlated spectroscopy case, it lowers faster with averaging compared to the classical case. When the measurement is not shot noise limited, the gain or loss using entanglement for spectroscopy depends on the details of the noise - the noise power spectral density. The reason is that when the frequency measured doubles, the noise amplitude arising from frequency shifts (e.g. magnetic field drifts on the ions) doubles.

%%%%%%%%%%%%%%%%%%%%%%%%%%%%%%%%%%%%%%%%%%%%%%%%%%%%%%%%%%%%%%%%%
% N ion generalization
%%%%%%%%%%%%%%%%%%%%%%%%%%%%%%%%%%%%%%%%%%%%%%%%%%%%%%%%%%%%%%%%%

In this work, only a two-ion crystal was used for a proof of principle experiment. However, the features demonstrated here are general and will apply for a larger number of spins as well. The required generalized Hamiltonian that will generate correlated $N$-spin rotations will be given by,
\begin{equation}
\begin{split}
\hbar\left[\Omega\left(\sigma_{x}\right)^{\otimes N}+\sum_{i=1}^{N}\delta_{i}I\otimes...\otimes\sigma_{z}^{i}\otimes...\otimes I\right].
\end{split}\label{Ising}
\end{equation}
Using this Hamiltonian, an N-fold narrower Rabi spectrum can be measured around the average resonance frequency. The simulation of the above $N$-body correlated Hamiltonian was proposed in \cite{Porras2009,Müller2011}. In principle, a universal quantum simulator can be used to implement multi-ion Heisenberg-limited Rabi spectroscopy on any number of spins.\\

%%%%%%%%%%%%%%%%%%%%%%%%%%%%%%%%%%%%%%%%%%%%%%%%%%%%%%%%%%%%%%%%%
%Conclusion
%%%%%%%%%%%%%%%%%%%%%%%%%%%%%%%%%%%%%%%%%%%%%%%%%%%%%%%%%%%%%%%%%
To conclude, in this work we presented and demonstrated a two-ions Heisenberg-limited Rabi spectroscopy. We initialized the ions in a separable state, and by operating with an entangling operator we obtained a spectrum narrower by a factor of $\simeq2$ with respect to conventional single ion Rabi spectroscopy. We observed that under the influence of an Ising Hamiltonian the two-ion system splits into two orthogonal subspaces that can be used as different probes for the difference and the average of the ions' optical resonance frequency, each of them with Heisenberg-limited uncertainty. We believe that the experiment presented here can be scaled up to more than a two-ion crystal, and may be useful as a spectroscopic tool for optical frequency measurements, as in optical atomic clocks.

%Acknowledgements:
This work was supported by the Crown Photonics Center, ICore-Israeli excellence center
circle of light, the Israeli Science Foundation, the Israeli Ministry of Science Technology
and Space, the Minerva Stiftung and the European Research Council (consolidator grant 616919-Ionology).

\pagebreak

\begin{center}
\textbf{\large Supplemental Materials}
\end{center}

\section{Section 1: light-shifting the resonance frequency of one ion }
In order to scan the frequency difference between our ions, we operated the single addressing beam on the desired ion, with frequency $3.5 \mathrm{MHz}$ off-resonance with the ion transition. This frequency shift was done with an  Acousto-Optic Deflector (AOD). The light intensity was varied by changing the AOD's RF source power, such that the light-shift achieved was between 0 and about $\simeq750 \mathrm{Hz}$. In order to calibrate the common frequency shift and the difference shift between the ions resulting from the light-shift, we performed uncorrelated Rabi spectroscopy on both ions using the global beam while the light-shifting beam was on. The results are shown in Figure 6.

\begin{figure}[H]
\begin{subfigure}[c]{1\textwidth}
\includegraphics[width=\textwidth]{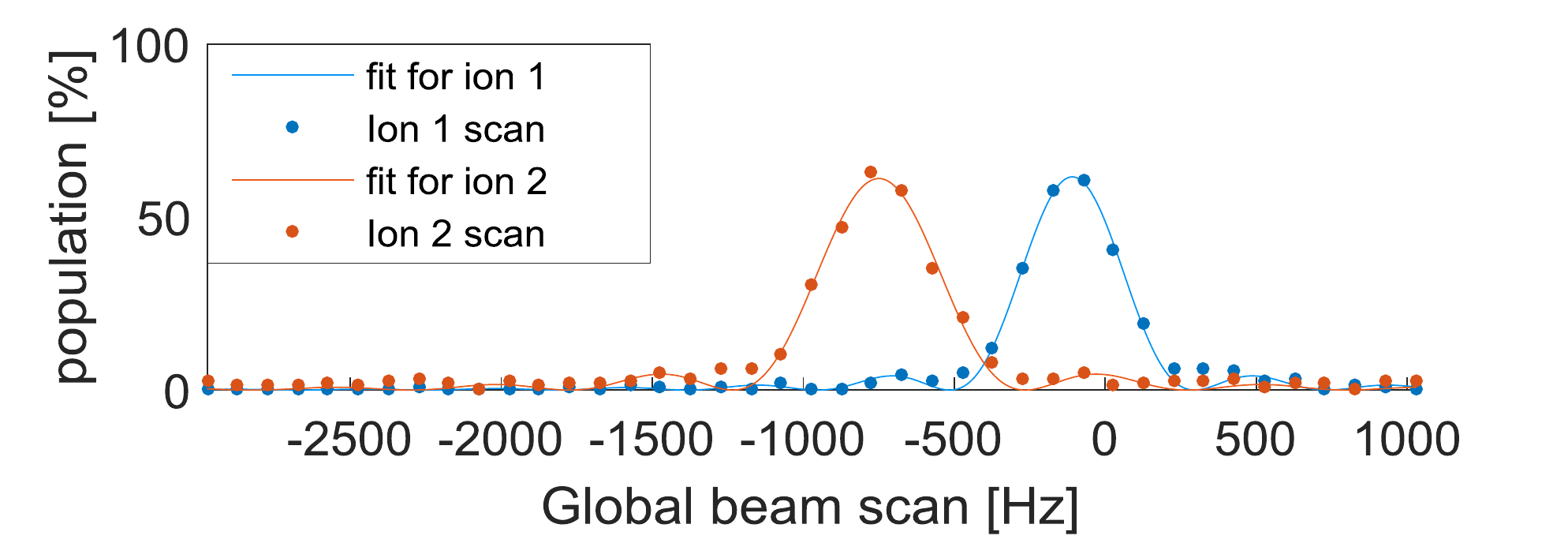}
\caption{}
\end{subfigure}
\hfill
\begin{subfigure}[t]{0.45\textwidth}
\centering
\includegraphics[width=\textwidth]{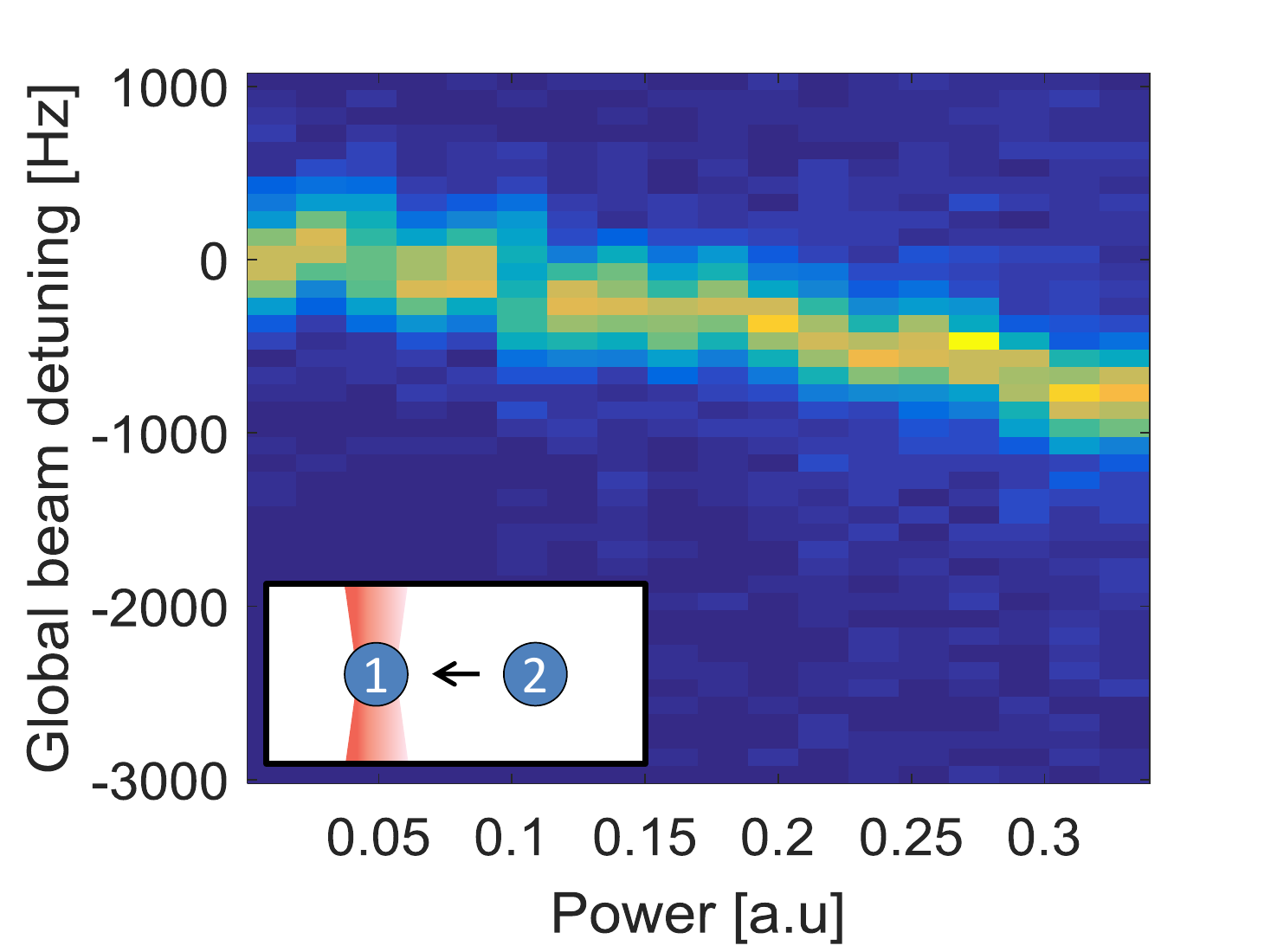}
\caption{}
\end{subfigure}
\hfill
\begin{subfigure}[t]{0.45\textwidth}
\centering
\includegraphics[width=\textwidth]{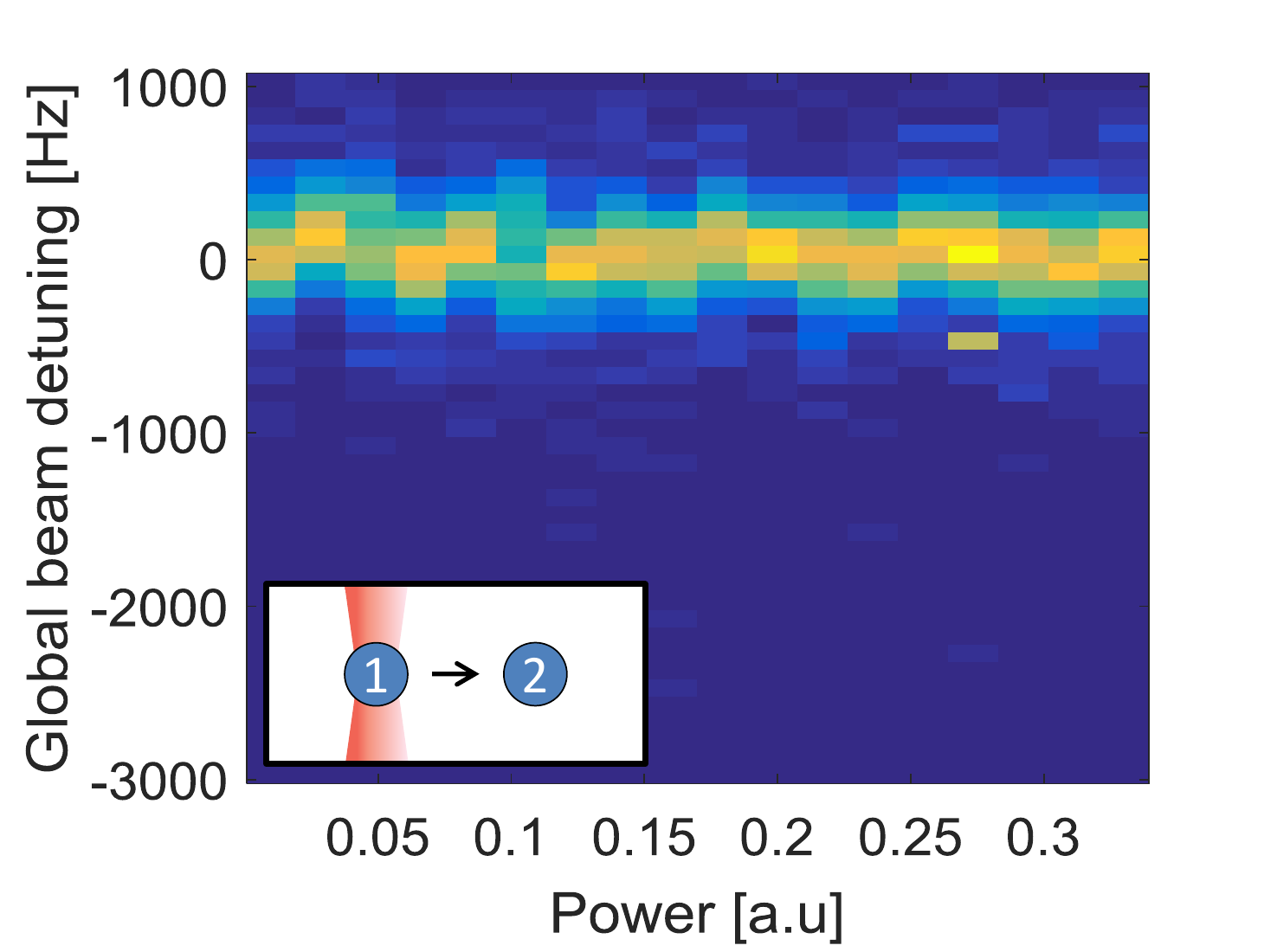}
\caption{}
\end{subfigure}
\hfill
\begin{subfigure}[t]{0.45\textwidth}
\centering
\includegraphics[width=\textwidth]{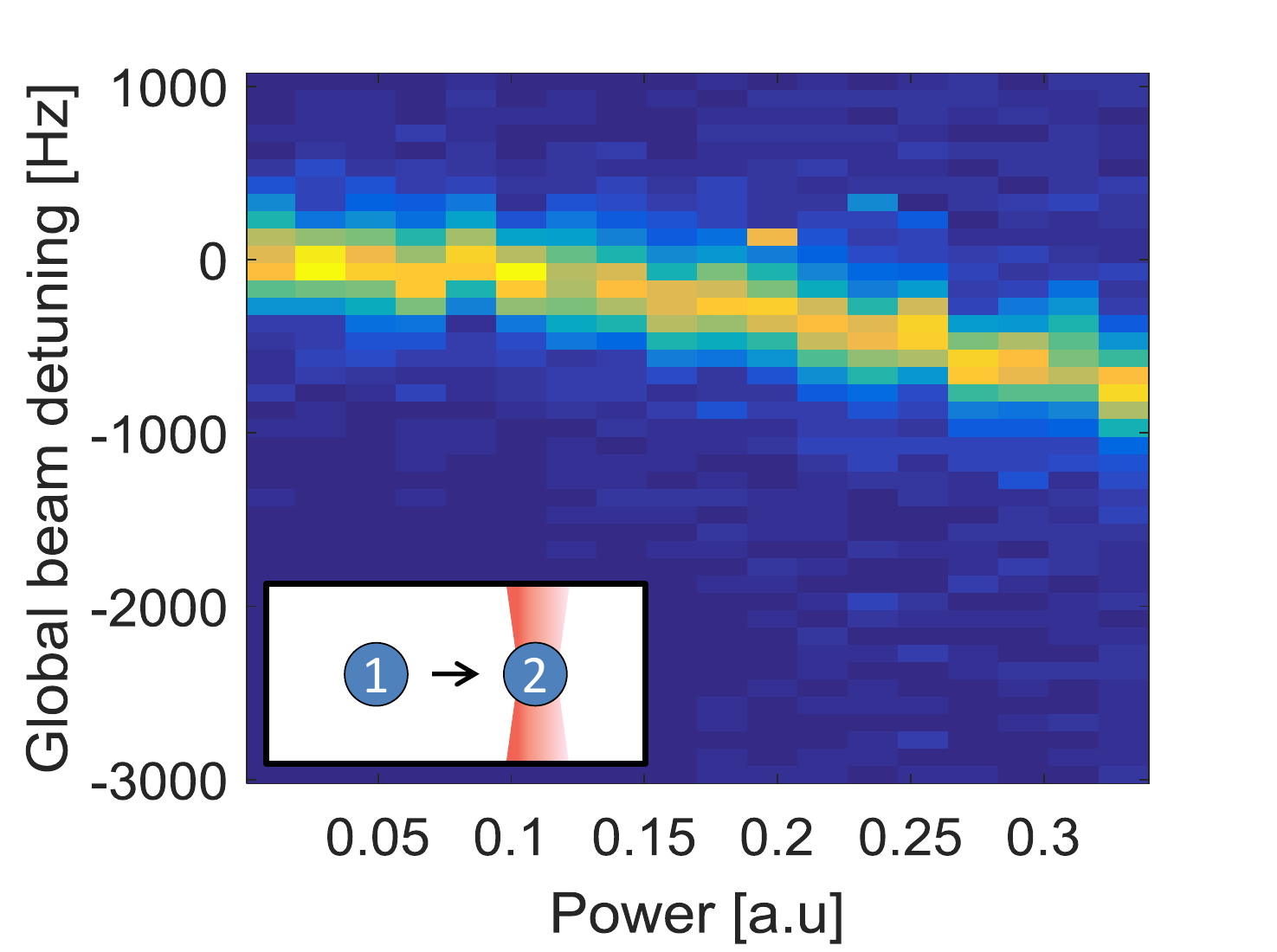}
\caption{}
\end{subfigure}
\hfill
\begin{subfigure}[t]{0.45\textwidth}
\includegraphics[width=\textwidth]{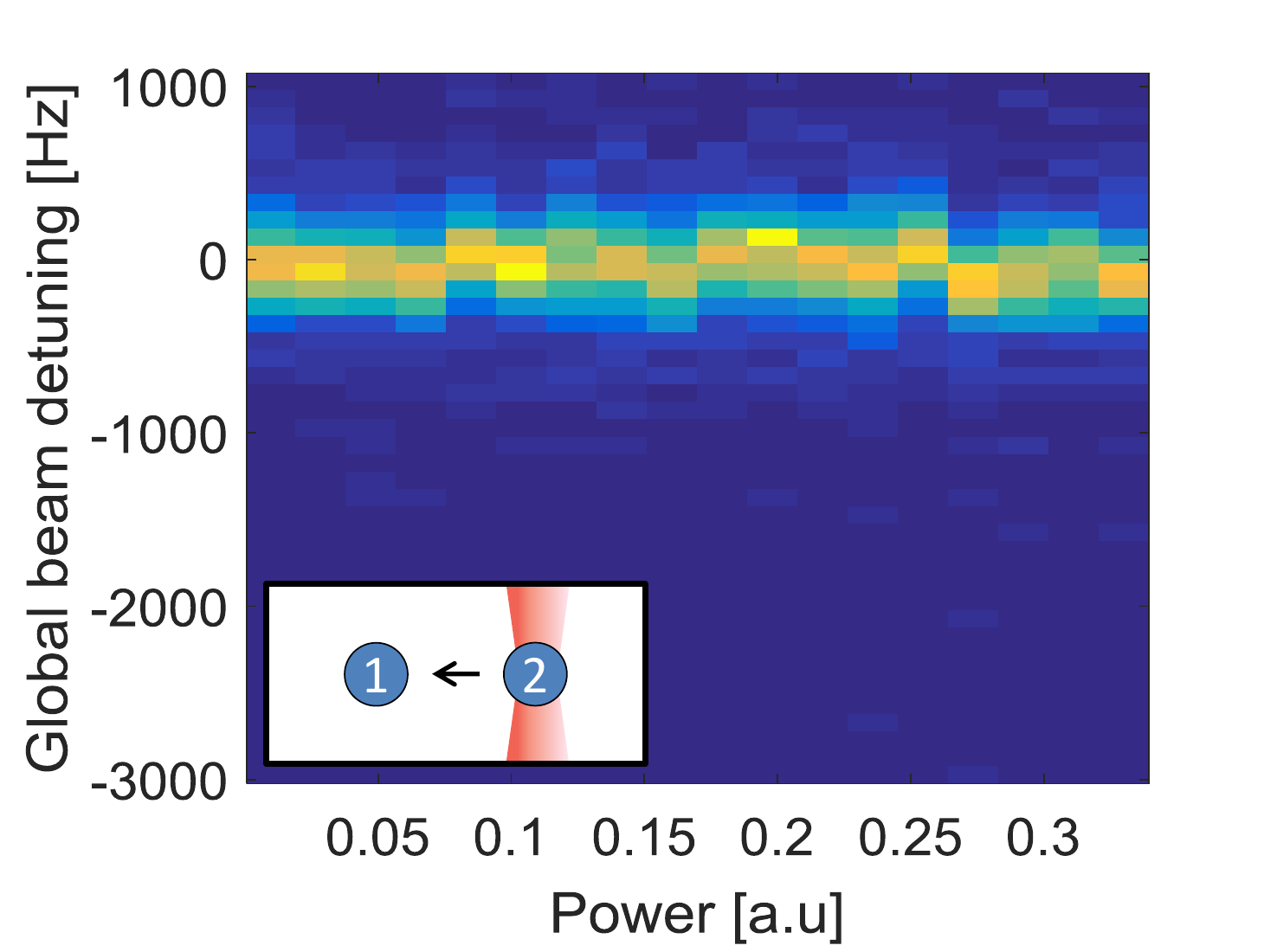}
\caption{}
\end{subfigure}
\hfill
\caption{\textbf{Measurement of light shift.} \textbf{(a)} A Rabi spectroscopy measurement example with light-shift beam on. \textbf{(b,c,d,e)} Rabi spectroscopy as a function of the light-shifting beam's power. The insets show which ion's spectrum is being taken (pointing arrow) and which ion is being light-shifted (red vertical beam). Zero global beam detuning was chosen as the averaged frequency between two non-light-shifted ions. }
\end{figure}

From these measurements we were able to calibrate the mean frequency of the ions and the difference between their frequency, as a function of the light-shift beam power. The calibration results are presented in Fig. 7.

\begin{figure}[H]
\includegraphics[width=\textwidth]{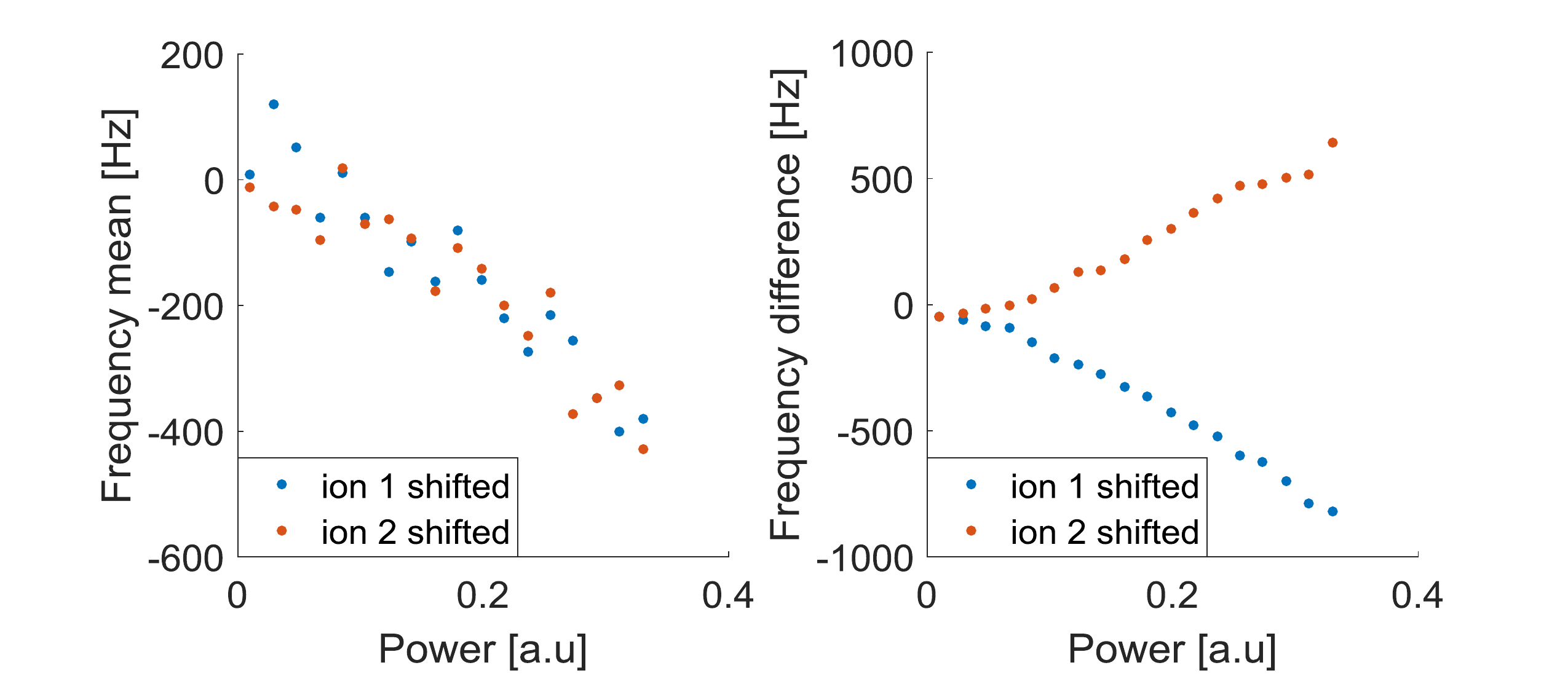}
\caption{\textbf{Calibration of common frequency shift and frequency difference between the two ions.} As predicted, mean frequency between the ions is the same for each of the ions being shifted, and the frequency difference between the ions changes its sign when moving from shifting one ion to the other. The frequency difference for zero light-shifting beam power is about 50 Hz, and we believe it is due to magnetic field gradient between the ions.}
\end{figure}

The values taken for the vertical axis of figure 4 are the values taken from the right plot in figure 7 divided by 2 (see figure 1f).\\
As a sanity check, we verified that indeed the mean-frequency shift measured from the Rabi spectroscopy results agrees with the shift of the correlated  $\left|\downarrow\downarrow\right\rangle\rightarrow\left|\uparrow\uparrow\right\rangle$ transition, that is apparent in figure 4a and 4d. The comparison is plotted in Fig 8.

\begin{figure}[H]
\centering
\includegraphics[width=0.55\textwidth]{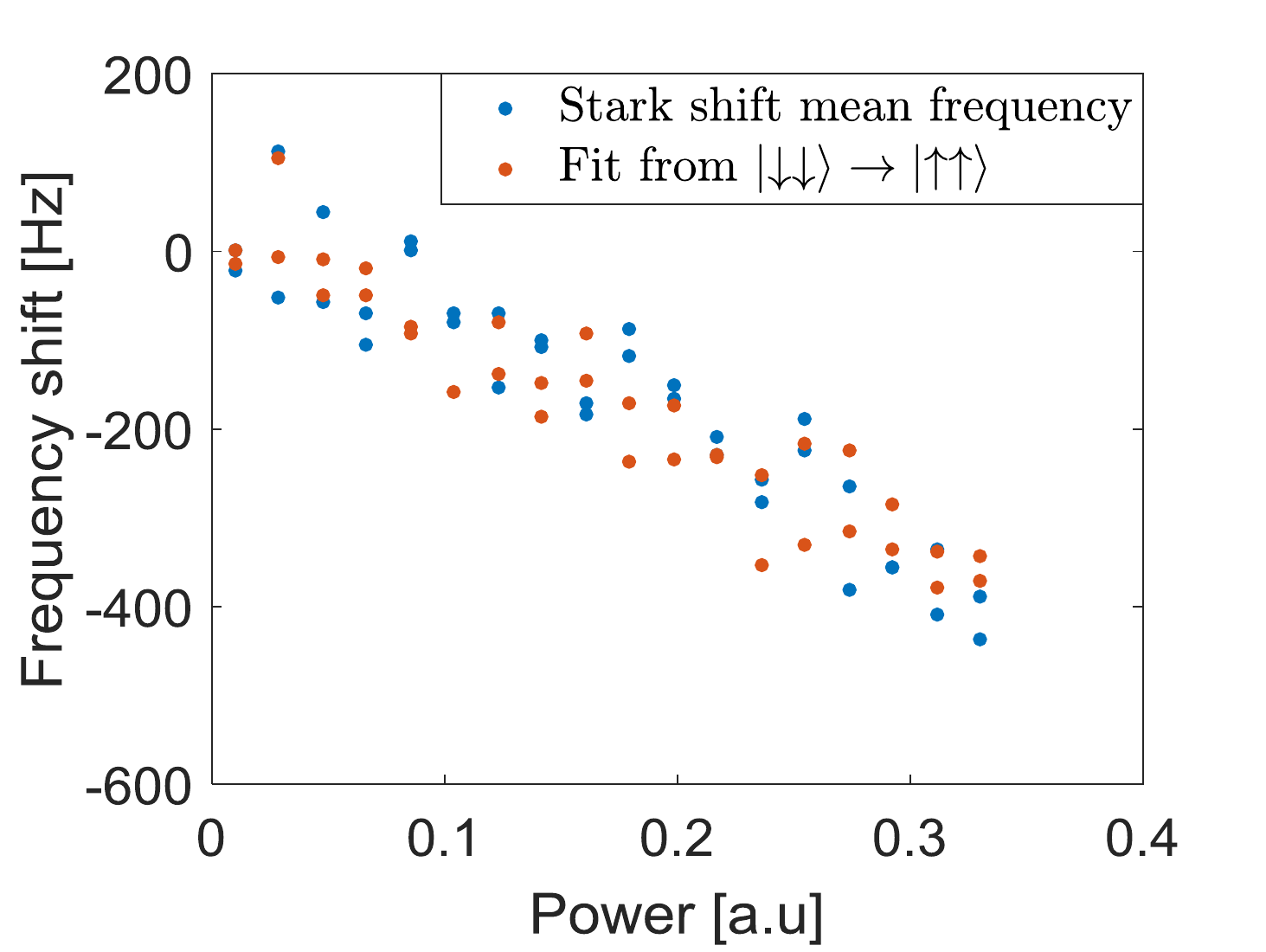}
\caption{\textbf{Comparison between resonance shift of correlated Rabi spectroscopy on $\left|\downarrow\downarrow\right\rangle\rightarrow\left|\uparrow\uparrow\right\rangle$ and mean frequency light shift measured on two uncorrelated ions.}}
\end{figure}

\section{Section 2: Frequency difference uncorrelated Rabi spectroscopy}
In figure 5b we present a comparison between a correlated and uncorrelated  frequency difference Rabi spectroscopy. The correlated part is described in the paper (see figure 1f). For this comparison we constructed a way to measure the frequency difference using two ions in an uncorrelated manner. We wanted to simulate a pure difference between the ions, without shifting of the frequency mean. Therefore, for each light-shift value we calibrated the mean frequency between the ions by performing a standard Rabi spectroscopy, fitting the spectrum and finding the mean frequency - $f_{mean}$. The experiment protocol was as follows: The ions were first initialized in the state $\left|\downarrow\downarrow\right\rangle$. Then a light-shifting beam was operated on one of the ions. Simultaneously, we operate a $\pi$ pulse with the global beam tuned to the $f_{mean}$. Lastly, a measurement of whether each ion is bright or dark was taken.\\

The measurement method is illustrated in Fig. 9.

\begin{figure}[H]
\centering
\includegraphics[width=0.55\textwidth]{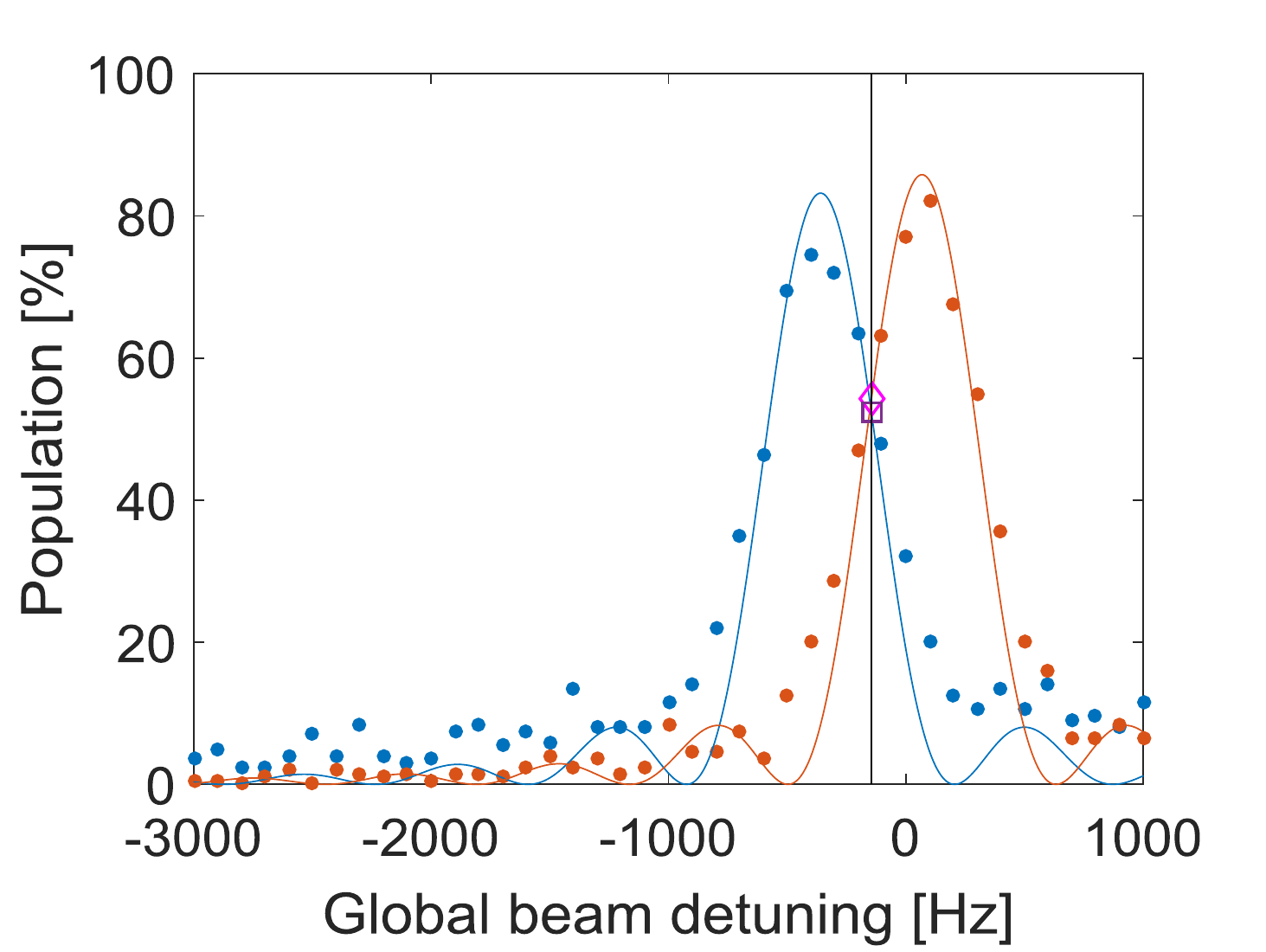}
\caption{\textbf{Measuring ion frequency difference.} The plot shows two single-ion Rabi spectroscopy spectra (red and blue full circles) where one ion frequency is light-shifted. The frequency mean was found using a fit (red and blue solid lines) to each ion's spectrum, and an additional measurement was taken at that point for each of the ions (hollow diamond and square)}
\end{figure}

\end{document}